\documentclass[structabstract]{aa}  
\usepackage{graphicx}
\usepackage[varg]{txfonts}
\usepackage{subfigure}
\usepackage{color}
\usepackage{natbib}
\bibpunct{(}{)}{;}{a}{}{,} 
\begin{document}

   \title{On the occurrence of galaxy harassment}
   \titlerunning{On the occurrence of galaxy harassment}
   \author{D. Bialas\inst{1}
          \and
           T. Lisker\inst{1}\fnmsep\thanks{\email{TL@x-astro.net}}
          \and
           C. Olczak\inst{1}\inst{2}\inst{3}
          \and
           R. Spurzem\inst{3}\inst{1}\inst{4}\inst{5}
          \and 
           R. Kotulla\inst{6}
          }
   \authorrunning{D.\ Bialas et al.}
   \institute{Astronomisches Rechen-Institut, Zentrum f\"ur Astronomie der Universit\"at Heidelberg, M\"onchhofstra\ss e 12-14, 69120 Heidelberg, Germany
         \and 
             Max-Planck-Institut f\"ur Astronomie (MPIA), K\"onigstuhl 17, 69117 Heidelberg, Germany         
         \and  
             National Astronomical Observatories of China and Key Laboratory for Computational Astrophysics, Chinese Academy of Sciences (NAOC/CAS), 20A Datun Lu, Chaoyang District, Beijing 100012, China
         \and 
             Kavli Institute for Astronomy and Astrophysics, Peking University, China
         \and
             Key Laboratory of Frontiers in Theoretical Physics, Institute of Theoretical Physics, Chinese Academy of Sciences, Beijing, 100190, China
         \and
             Center for Gravitation and Cosmology, Department of Physics, University of Wisconsin - Milwaukee, 1900 E Kenwood Blvd., Milwaukee WI, 53211, USA
             }
   \date{Accepted 12 February; 2015 Received 29 October 2014}
   
 
  \abstract
{Tidal interactions of galaxies in galaxy clusters have been proposed as one potential explanation of the morphology-density relation at low masses. Earlier studies have shown that galaxy harassment is a suitable mechanism for inducing a morphological transformation from low-mass late-type disk galaxies to the abundant early-type galaxies.}
{The efficiency of tidal transformation is expected to depend strongly on the orbit of a galaxy within the cluster halo. The orbit determines both the strength of the cluster's global tidal field and the probability of encounters with other cluster members. Here we aim to explore these dependencies.}
{We use a combination of N-body simulation and Monte Carlo method to study the efficiency of the transformation of late-type galaxies by tidal interactions on different orbits in a galaxy cluster. Additionally, we investigate the effect of an inclination between the disk of the infalling galaxy and its orbital plane. We compare our results to observational data to assess the possible relevance of such transformations for the existing cluster galaxy population.}
{We find that galaxies that entered a cluster from the outskirts are unlikely to be significantly transformed (stellar mass loss $\leq$ 6\%). Closer to the cluster centre, tidal interactions are a more efficient mechanism (stellar mass loss up to 50\%) for producing harassed galaxies. The inclination of the disk can reduce the mass loss significantly, yet it amplifies the thickening of the galaxy disk. Galaxies with smaller sizes on intermediate orbits are nearly unaffected by tidal interactions.
The tidal influence on an infalling galaxy and the likelihood that it leads to galaxy harassment make a very stochastical process that depends on the galaxy's specific history.}
{We conclude that harassment is a suitable mechanism that could explain the transformation of at least a fraction of galaxies inside galaxy clusters. However, the transformation would have to start at an early epoch in protocluster environments and continue until today, in order to result in a complete morphological transformation.}
 
  
   \maketitle
%

\section{Introduction}

For bright galaxies, \cite{1980ApJ...236..351D} showed that the population fraction of elliptical galaxies in clusters increases with increasing local environmental galaxy density, while at the same time the fraction of spiral galaxies decreases. 
An analogous morphology-density relation is observed for dwarf galaxy types \citep{1987AJ.....94..251B}. Dwarf irregulars and low-mass spirals dominate outside of dense environments, while early-type dwarf galaxies (dE, subsuming objects termed dwarf lenticulars) are the most common type of galaxy in clusters.
However, the origin of this large population of dEs is still a topic of ongoing debate,
partly because they were found to follow a number of continuous relations in structure and colour with bright elliptical and/or lenticular galaxies \citep[e.g.][]{Cote2007,Chen2010,Kormendy2012}, but
 also since their identification and detailed analysis are not yet possible at higher shifts (see \citet{2009A&A...508..665B} for dEs in a z=0.165 cluster).
%
%
%
Clues to the origin of dEs may be given by the presence of weak disk features in a significant fraction of them \citep{2000A&A...358..845J,DeRijcke2003,2006AJ....132..497L}, the presence of residual star formation activity in dEs outside of the densest environments \citep{2006AJ....132.2432L,2008AJ....135.1488T,Pak2014} and the finding that these have significantly flatter shapes than dEs  with low relative velocities in the Virgo cluster core \citep{2009ApJ...706L.124L}. Taken together, this may be regarded as supporting the hypothesis that many dEs were formed from disk galaxies that were affected and eventually transformed by a dense environment \citep[for further observational aspects related to this hypothesis, see also][]{Kormendy1985,Binggeli1990,DeRijcke2005,Beasley2006,SmithRussell2012,Rys2014}.
In this scenario, dEs with disk features (dEdis) could be galaxies in an incomplete transformation phase \citep{GrahamJerjenGuzman2003,2005MNRAS.364..607M}. 

Different processes have been discussed to explain the morphological transformation. In general, they can be divided into two main types based on the underlying physics: hydrodynamical scenarios and tidal scenarios. In reality, a mix of both types will act on an infalling galaxy, but it is still an open and relevant question how strong the impact is for which kind of interaction. One of the most popular tidal scenarios is galaxy harassment \cite{1996Natur.379..613M}.  
It is characterised as the combined effect of tidal interactions between an infalling galaxy and the global cluster potential and of the tidal interaction with close high speed encounters with other cluster members. \cite{1996Natur.379..613M} used a set of N-body simulations to show that these tidal interactions are suitable for causing a morphological transformation of spiral galaxies. This outcome is generally supported by studies of various authors, even though debating the actual efficiency \citep{1998ApJ...495..139M, 1999MNRAS.304..465M, 2005MNRAS.364..607M, 2009A&A...494..891A, 2010MNRAS.405.1723S}.

The efficiency of the harassment scenario is expected to depend on the orbit of an infalling galaxy within a galaxy cluster \citep{2010MNRAS.405.1723S}. The deeper an orbit is placed inside a cluster, the stronger the tidal force of the background mass of the cluster and the higher the probability of close encounters with other cluster members. 
\cite{2005MNRAS.364..607M} show in their simulations of bulgeless disk galaxies with masses of $7 \cdot 10^{10} M_{\odot}$ that combined tidal interactions are efficient at harassing galaxies in the innermost regions of a cluster on orbits with pericentres very close to the cluster centre (between 40kpc and 100kpc) and less efficient on orbits in the outer regions of a cluster. \cite{2010MNRAS.405.1723S} show that late-type dwarf irregular galaxies with masses of $10^{10} M_{\odot}$ that enter a cluster from its virial radius are only slightly affected and not harassed, but they find that dwarfs on orbits very close to the cluster centre comparable to those of \cite{2005MNRAS.364..607M} could be harassed efficiently. Both studies have shown what happens to late-type galaxies of different mass ranges on orbits very close to the cluster centre and far away from it. But especially on more moderate orbits, the question is still open whether galaxies could be efficiently harassed or not.
With the study presented here, we intend to add a piece to that puzzle.  

Complementary to the orbital parameters, \cite{1999MNRAS.304..465M} show that the properties of the progenitor galaxies are relevant for the efficiency of  transformation by tidal interactions. They show that compact high surface brightness galaxies (HSB) are hard to affect, and more extended low surface brightness galaxies (LSB) can be affected efficiently. A comparative result was shown by \cite{2013MNRAS.431.3533C} by probing the efficiency of tidal stripping for different morphologies of satellite galaxies in host galaxy of Milky Way-like mass. 
In their simulations of disk galaxies orbiting inside dark matter haloes with group masses, \cite{2012MNRAS.424.2401V} have additionally found that the inclination between the orbital plane and the disk of the infalling galaxy is an important parameter for the effect of tidal stripping. However, tidal stripping is only the most extreme effect of tidal interactions, and one has to expect that an inclination of the infalling galaxy may affect the reshaping of the galaxy. 

To sum up, the parameter space for the efficiency of the tidally induced transformation in a galaxy cluster spans the orbital parameters, the type of progenitor, and the orientation of the galactic disk. Furthermore, the properties of the cluster itself, such as its mass, its galaxy density, and the velocity dispersion of its galaxies, are expected to be relevant. Some parameters are partly degenerated; e.g.,\ a variation in the orbit and the cluster properties could end up with a similar tidal strength.  
To determine the contribution of harassed galaxies to the population of dEs in clusters, it is necessary to understand the efficiency of harassment depending on these parameters.
 
In this study, we investigate the efficiency of tidally induced transformations in a Virgo-like cluster by varying the orbital parameters. We adopt three different orbits. On the innermost orbit, we use three different orientations of the disk with respect to the orbital plane. Furthermore, on the innermost orbit, we probe how the efficiency changes if the progenitor galaxy is less extended. 
We use a simple stellar evolution approach to compare our galaxy remnants to the observed dE population of the Virgo cluster. 
The aim of this study is to show the relevance of the parameters, as well as their general trends with respect to transformation efficiency. In section 2 we outline the numerical method and the different setups. Results of the different simulations are presented in section 3. In section 4 we compare our simulations with observations. In section 5 the results are discussed and interpreted, leading to our conclusions in section 6.
  

\section{Numerical methods}

There are many parameters that affect the tidal interaction of an infalling galaxy inside a galaxy cluster: the orbital parameters, the structural parameters of the infalling galaxy, and the structural parameters of the cluster. For this study we decided to keep the cluster fixed and concentrate mostly on the orbital parameters, but including a variation of the structural parameters of the infalling galaxy. 
We used a combination of N-body simulations and a Monte-Carlo method to simulate the harassment scenario.
For these simulations the direct $N$-body integrator NBODY6 was taken \citep{1999PASP..111.1333A,Aarseth2003}, with GPU acceleration by \citet{2012MNRAS.424..545N} and the massively parallel version using MPI \citep{1999JCoAM.109..407S,Spurzem2008}. This code, called NBODY6++GPU (Wang et al., in prep.), was modified to treat strong and time-varying tidal fields (see Appendix) and to simulate encounters with massive point-like particles for the harassment scenario. 

While the current version of the code can simulate up to $10^6$ particles now on a few dozen GPU nodes (order $10^5$ GPU cores for Kepler GPUs), we used only much more moderate particle numbers; any unwanted two-body relaxation occurring due to too small a particle number has been corrected for (see Sect.~2.3). The code contained the SSE (single-stellar evolution) package of \citet{2000MNRAS.315..543H} and has been combined with the GALEV spectral synthesis code (\citealt{2009MNRAS.396..462K}; Pang et al., in prep.). Although NBODY6 and its offsprings were originally designed for star-by-star cluster simulations, the code turned out to be very useful in this project, also thanks to the GPU acceleration. We have not resolved our galaxies star-by-star, so a single body in the simulation typically represents many stars. 

\subsection{Cluster model}

The gravitational force of the cluster was modelled as an analytical force field. An Navarro–Frenk–White profile \citep{1996ApJ...462..563N}  was adopted to describe the density profile of the galaxy cluster, and the fit parameters $\rho_0 = 3.2 \cdot 10^{-4} M_{\odot}/pc^{-3} $ and $ R_s $ = 560  kpc  from \cite{1999ApJ...512L...9M}  were used to reproduce the force field of a Virgo-like cluster  with $ M_{200} = 4.2 \cdot 10^{14} M_{\odot}$  and \mbox{ $R_{200}$ = 1.55 Mpc}. Nbody6++GPU is using Hill's approximation to take external force-fields into account. In this approximation it is assumed that the galaxy is moving on an circular orbit around the cluster centre, which allows the Euler forces to be neglected. We have added some code to Nbody6++GPU to include the Euler forces, so that we are able to use eccentric orbits for the galaxies around the cluster centre. An explanation of the equations of motions that are used is given in Appendix A.
The orbit of the infalling galaxy around the cluster centre was calculated using a leap-frog algorithm. Free parameters of the orbit were the apocentre and pericentre; we kept the pericentre fixed at 0.3 Mpc and used three different apocentres (0.5 Mpc, 1.0 Mpc, and 1.5 Mpc). 

\subsection{Encounters with other cluster members}
 
Flyby events of cluster members were simulated by using a Monte-Carlo method. A random number test was used to check periodically if an encounter happened or not. In the case of an encounter, an additional particle was injected into the simulation, which represents the flying-by cluster member. After that we waited half of the typical time between two encounter events and checked again whether another encounter had happened. 

The typical time between two encounters depends on the local galaxy density and the local velocity dispersion of the cluster. In this way the encounters with other cluster members are more frequent near the cluster centre. The local galaxy density was derived from the radial galaxy distribution of those galaxy clusters of the Millennium II simulation \citep{2009MNRAS.398.1150B} that have masses between $2.4 - 4 \cdot 10^{14} M_{\odot}$. We made use of the corresponding semi-analytic model of \cite{2011MNRAS.413..101G} to select cluster member galaxies with total (dark+stellar) masses of more than 0.1 $M_{\rm gal}$, where $M_{\rm gal}$ is the initial total mass of our simulated infalling galaxy.
The local velocity dispersion was calculated numerically as a solution of the spherical Jeans-equation for the cluster potential.   

The masses and velocities of the injected particles were chosen randomly. The masses following the  mass function of the aforementioned galaxy clusters of the Millennium II simulation \citep{2009MNRAS.398.1150B}. The minimum mass of a flying-by galaxy that was assumed to be relevant strong enough to disturb the infalling galaxy,  hence is worth considering, was $0.1 M_{ \rm gal}$. The velocities follow a Maxwell-Boltzmann distribution, and the velocity orientation is chosen randomly. 

The impact parameters of the encounters were chosen by following the geometrical cross section. Two different galaxy models with different extents were used in this study. The minimum impact parameter for simulations with the more extended galaxy model was set to 8 kpc and for simulations with the smaller galaxy model to 4 kpc, the maximal impact parameter was set to 60 kpc. Owing to gravitational focusing, the distance of minimal approach will be closer than the impact parameter at infinity. To take this into account and to correct the impact parameter at infinity, we enlarged it by using the relation of the closest approach and impact parameter of \cite{2009ApJ...697..458S}. A more detailed description on how the parameters of encounters were chosen is given in Appendix B.

\subsection{Galaxy models}

The galaxy models were three component models with a stellar disk and bulge and a dark matter halo created by the GalactICS code of \cite{1995MNRAS.277.1341K}. 
This code uses a King model \citep{1962AJ.....67..471K} for the bulge, an exponential Shu model \citep{1969ApJ...158..505S} for the disk, and an Evans model \citep{1993MNRAS.260..191E} for the dark matter halo. 

Two different galaxy models were used, one with a disk scale length $r_0$ of 3 kpc and one with a disk scale length of 1.2kpc. The 3kpc model was our main model, which was used in 20 of 24 simulations and is referred to as the "large model". The 1.2 kpc model is referred to  as the "small model". 
The large model is based on the typical values of dust-corrected k-band observations of Sc/Sd galaxies \citep{2008MNRAS.388.1708G}, which we assume to be suitable late-type progenitor galaxies. The mass-to-light ratios for the conversion of the k-band values were 0.46 for the disk and 0.82 for the bulge component. We determine these conversion factors with the GALEV evolutionary synthesis models from \cite{2009MNRAS.396..462K} with Sandage-functions \citep{1986A&A...161...89S} for the star formation histories and delay times of $\tau$ = 3Gyr for the bulge and $\tau$ = 9Gyr for the disk. 

The small model is constructed to be a dwarf version of an Sc/Sd galaxy. The properties of both are listed in Table~\ref{table:1}. 
With a thickness of $ r_{0}/z_{0} = 9.2 $ for the small model and $ r_{0}/z_{0} = 6.7 $ for the large model, both models have a comparatively thin disk component \citep{2009ApJ...702.1567B}. 
The models were represented by $10^{5}$ particles, with $3 \cdot 10^{4}$ disk particles, $5 \cdot 10^{3}$ bulge particles, and $6.5 \cdot 10^{4}$ halo particles. 
Because Nbody6++GPU is a collisional N-Body code, one needs to correct the simulation results with respect to two-body relaxation. To determine the evolution under two-body relaxation, a control simulation was made for each of both models. In this control simulation, the evolution of the galaxy models without any tidal field or galaxy encounters was simulated, yielding the evolution of the model galaxy parameters that is caused by two-body relaxation alone. Following the idea of \cite{1999MNRAS.304..254V}, one can correct the measured parameters from the simulations by subtracting the reference values of the parameter change caused by two-body relaxation, as determined by the control simulation. In that way one gets the relaxation-free change of the model parameters. We have cross-checked this assumption, by simulating two of our simulations again, doubling the total number of particles, and it turned out that the correction works fine. The differences in the relaxation-free values were in the range $5\% - 10\%$.

%
\begin{table}
\caption{Galaxy models}             
\label{table:1}      
\centering                          
\begin{tabular}{c c c c c c }        
\hline\hline                 
Model & $M_{\rm tot}$ & $M_{\rm star}$ & D/B & $r_{0}$ & $z_{0}$  \\    
\hline                   
small    & 1.74 & 0.12 & 6.4 &  1.2 & 0.13 \\
large    & 5.00 & 0.35 & 6.0 &  3.0 & 0.45 \\
\hline                                   
\end{tabular}
\tablefoot{ (1) Total mass in $10^{10} M_{\odot}$; (2) Stellar mass in $10^{10} M_{\odot}$; (3) Disk to bulge ratio; (4) Scale length of the disk in kpc; (5) Scale height of the disk in kpc}
\end{table}

\subsection{Setups}

\begin{figure}
   \includegraphics[angle=-90. , width=0.5\textwidth]{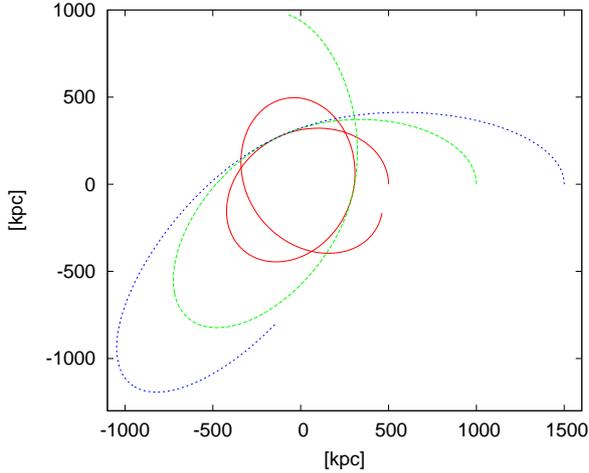}
      \caption{Orbits with a pericentre of 0.3 Mpc and an apocentre of 0.5 Mpc (red), 1.0 Mpc (green dashed), and 1.5 Mpc (blue dotted).}
         \label{Orbit-Plot}
   \end{figure}

Three different orbits were used in this study: their apocentres were \mbox{1.5 Mpc}, \mbox{1.0 Mpc,} and \mbox{0.5 Mpc,} and all of them have the same pericentre at \mbox{0.3 Mpc}. The different orbits are shown in Fig.~\ref{Orbit-Plot}. All simulations started at their apocentre and spanned a time of 5 Gyr. The time span of 5 Gyr is similar to previous studies \citep{1996Natur.379..613M, 1998ApJ...495..139M, 1999MNRAS.304..465M, 2005MNRAS.364..607M}, and it was chosen to follow the evolution of galaxies inside a cluster environment since z $\simeq$ 0.5 . 

To probe the importance of the initial structural parameters of a galaxy, we ran one simulation setup with a  smaller galaxy model.  
It was shown by \cite{2012MNRAS.424.2401V} that the inclination of the disk to the orbital plane could change the evolution of a galaxy in a group or cluster environment, in the way that galaxies with an inclination could resist the tidal forces for a longer time. To investigate how strong this impact is, we took three different inclinations for the galaxies on the innermost orbit: the disk plane parallel to the orbital plane $0^\circ$, the disk plane perpendicular to the orbital plane $90^\circ$, and one orientation between $45^\circ$. 

We made three simulations per setup, each with a different random number seed to get different realisations of the random processes. In addition, one simulation per setup was made without encounters with other cluster members, to study the influence of the tidal field alone and to investigate the stochasticity of the harassment scenario. Our setups are listed in Table~\ref{table:2}.

%
\begin{table}
\caption{Simulation setups}             
\label{table:2}      
\centering                          
\begin{tabular}{c c c c c c}        
\hline\hline                 
 Model & Apo & Peri & Inc & $M_{\rm tot}(t=0)$\\    
\hline                   
 large & 0.5 & 0.3 & 0 & 37 \\ 
 large & 1.0 & 0.3 & 0 & 57 \\
 large & 1.5 & 0.3 & 0 & 73 \\
 small  & 0.5 & 0.3 & 0 & 100 \\
 large & 0.5 & 0.3 & 45 & 37 \\
 large & 0.5 & 0.3 & 90 & 37 \\
\hline                                   
\end{tabular}
\tablefoot{ (1) Used galaxy model; (2) apocentre in Mpc; (3) pericentre in Mpc; (4) inclination in dgre; (5) initially bound total mass in \%}
\end{table}
%
%

   \begin{figure}
   \centering
   \includegraphics[angle=-90. ,width=0.5\textwidth]{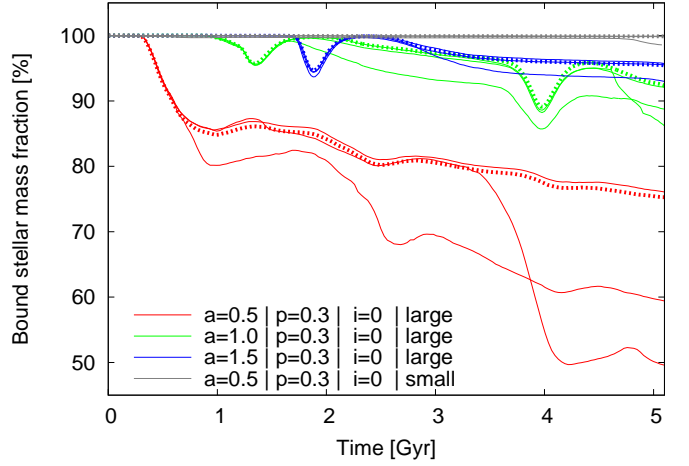} 
   \caption{Time evolution of the bound stellar mass fraction in simulations with different orbital parameters of the large galaxy model (coloured) and with the small model (grey).
The dotted lines represent simulations without encounters with other cluster members.}
   \label{MstarOrb}
   \end{figure}

\subsection{Bound mass}

 The following algorithm was used to determine the bound particles.  We calculated the kinetic and potential energy of each particle and checked whether the absolute value of the potential energy was higher than its kinetic energy. In that case the particle was bound energetically, but it remained unclear whether it was bound to the galaxy or only to the cluster. Particles that were outside the Lagrangian points L1 and L2 of the galaxy are not bound to the gravitational centre of the galaxy but to the cluster and will leave the galaxy following the attraction of the cluster. To distinguish between both cases, we used the tidal radius $R_T$, which is an approximation of the distance from the galaxy centre to the Lagrangian points L1 and L2, in order to clarify whether a particle lies beyond the Lagrangian points or not. Because the tidal radius itself is only an approximation of this problem, we used the criteria for the discrimination that a particle has to be inside of $1.5 R_T$ around the density centre of the bulge,in order to be bound to the galaxy.  

All simulations started at the apocentre of their orbits. At this point the initial tidal radius was mostly already smaller than the extent of the dark matter halo and therefore only part of the whole mass of the galaxy was bound initially. The last column of Table~\ref{table:2} shows the initially bound mass fractions.

\subsection{Virtual photometry}

The photometric properties of different galaxy types are not randomly distributed in phase space, but the different galaxy types populate specific regions of the phase space. 
If the tidal transformation of late-type galaxies by the harassment scenario is one of the main production channels of dE galaxies in galaxy clusters, the remnants of harassed galaxies have to match the photometric properties of the dE galaxies. 
To investigate whether the simulated remnants fulfil this condition, the photometric properties of the remnants were calculated and compared to those of dE galaxies in the Virgo galaxy cluster. 

To compute the brightness and the half-light radius of the simulated galaxies, the stellar evolution of the bulge and disk component 
were simulated by using the single stellar evolution code of \cite{2000MNRAS.315..543H}. To do so, we adopted a Sandage-function \citep{1986A&A...161...89S} for the star formation history, following \cite{2002ApJ...576..135G}, with a delay time $\tau$ = 3Gyr for the bulge and $\tau$ = 9Gyr for the disk. As initial mass function, the Kroupa IMF \citep{2001MNRAS.322..231K} in a mass range of $0.08M_{\odot} - 8M_{\odot}$ was used. The metallicities of the stars were calculated by using the chemically consistent evolution of the GALEV population synthesis models \citep{2009MNRAS.396..462K}. Each stellar particle was assigned an individual mass, age, and metallicity, and the evolution of the star was traced. Finally parts of the GALEV population-synthesis code \citep{2009MNRAS.396..462K} were used to convert the stellar properties into the r-band brightness of the stars by using the BaSeL-library \citep{1997A&AS..125..229L, 1998A&AS..130...65L}.

Furthermore, it was assumed that the star formation ended abruptly when the galaxy entered the cluster. While it is expected that a low-mass galaxy loses most of its gas relatively quickly by ram pressure \citep{1972ApJ...176....1G, 2008ApJ...674..742B} -- even outside the virial radius when entering along filaments \citep{2013MNRAS.430.3017B} -- this is certainly a simplified approach. Models employing an instantaneous removal of the extended gas reservoir of galaxies that become satellites led to an overabundance of passive galaxies in clusters \citep{2010MNRAS.406.2249W}; a recent study combining observations and simulations instead points to a picture in which ''satellite quenching is delayed-then-rapid'' \citep{2013MNRAS.432..336W}. On the other hand, the relevance of pre-processing \citep{2012MNRAS.424.2401V} may imply an earlier decline of the star formation rate. Since we are primarily interested in simulating the effects on the stellar structure, we consider the star formation rate to be zero from the starting point of our
simulation, 5 Gyr ago. 

To compute the total brightness and the half-light radius of the galaxies, projections of the galaxies were created under different projection angles  $\Theta \in [0^{\circ}, 22.5^{\circ}, 45^{\circ}, 67.5^{\circ}, 90^{\circ}]$ along the x and y-axes. We extracted radial profiles by fitting ellipses to the isophotes following the formula of \cite{1996A&AS..117..393B}.  As total brightness of the galaxy, the brightness inside of an elliptical aperture of two Petrosian semi-major axes was used \citep{1976ApJ...209L...1P,2007ApJ...660.1186L}.
The major axis of the elliptical profile which encloses half of the total luminosity was adopted as the half-light radius of the galaxy. From the total brightness and the half-light radius one can compute the effective surface brightness.

For the comparison with observations, data from the Sloan Digital Sky Survey (SDSS) \citep{2006ApJS..162...38A} were used. The sample consists of 864 members and possible members of the Virgo cluster, classified as dE, dEdi, S0, Sc, Sd, Sm, and Im galaxies  based on the Virgo Cluster Catalogue \citep{1985AJ.....90.1681B} down to the completeness limit of a B-band apparent magnitude 18 and with velocities below 3500 km/s. The measurements of r-band brightness, surface brightness, and half-light radius were taken from \cite{2007ApJ...660.1186L},  \cite{2008ApJ...689L..25J, 2009AN....330..948J}, and \cite{2014A&A...562A..49M}. The sample comprises 449 dE , 41 dEdi, 73 S0, 114 Sc/Sd, 26 Sm, and 161 Im galaxies.

\section{Results}


In the following section, we want to analyse the impact of orbital parameters, the structural parameters of a galaxy, and the inclination of a galaxy. In Table~\ref{table:3}, the initial properties of the galaxy models, as well as the properties of the galaxy remnants of the different simulations, are listed. The values of Table~\ref{table:3} are already relaxation-corrected as explained in section 2.3.

   \begin{figure*}
   \subfigure{ 
           \includegraphics[angle=-90. ,width=0.363\textwidth]{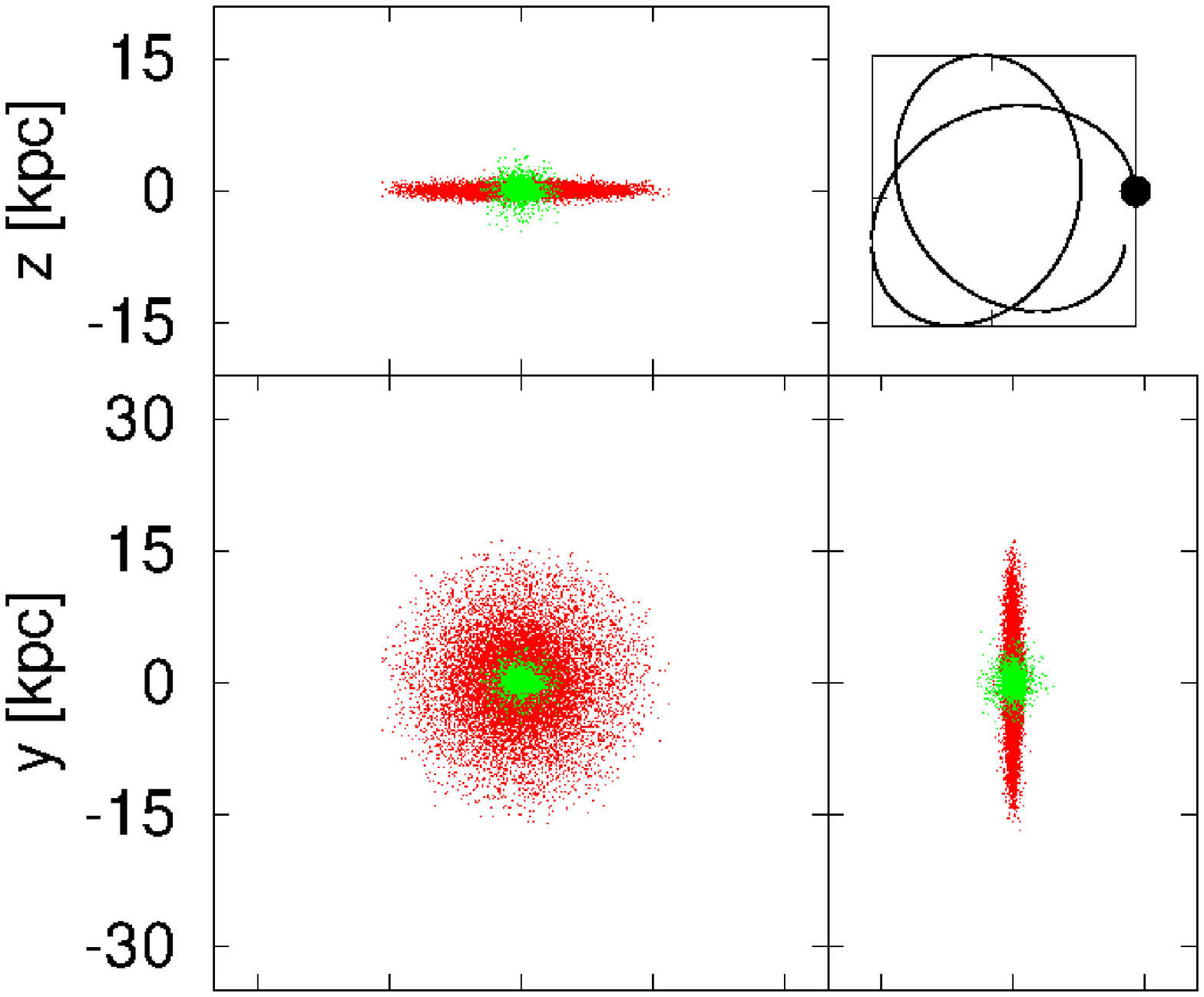}
           \includegraphics[angle=-90. ,width=0.3\textwidth]{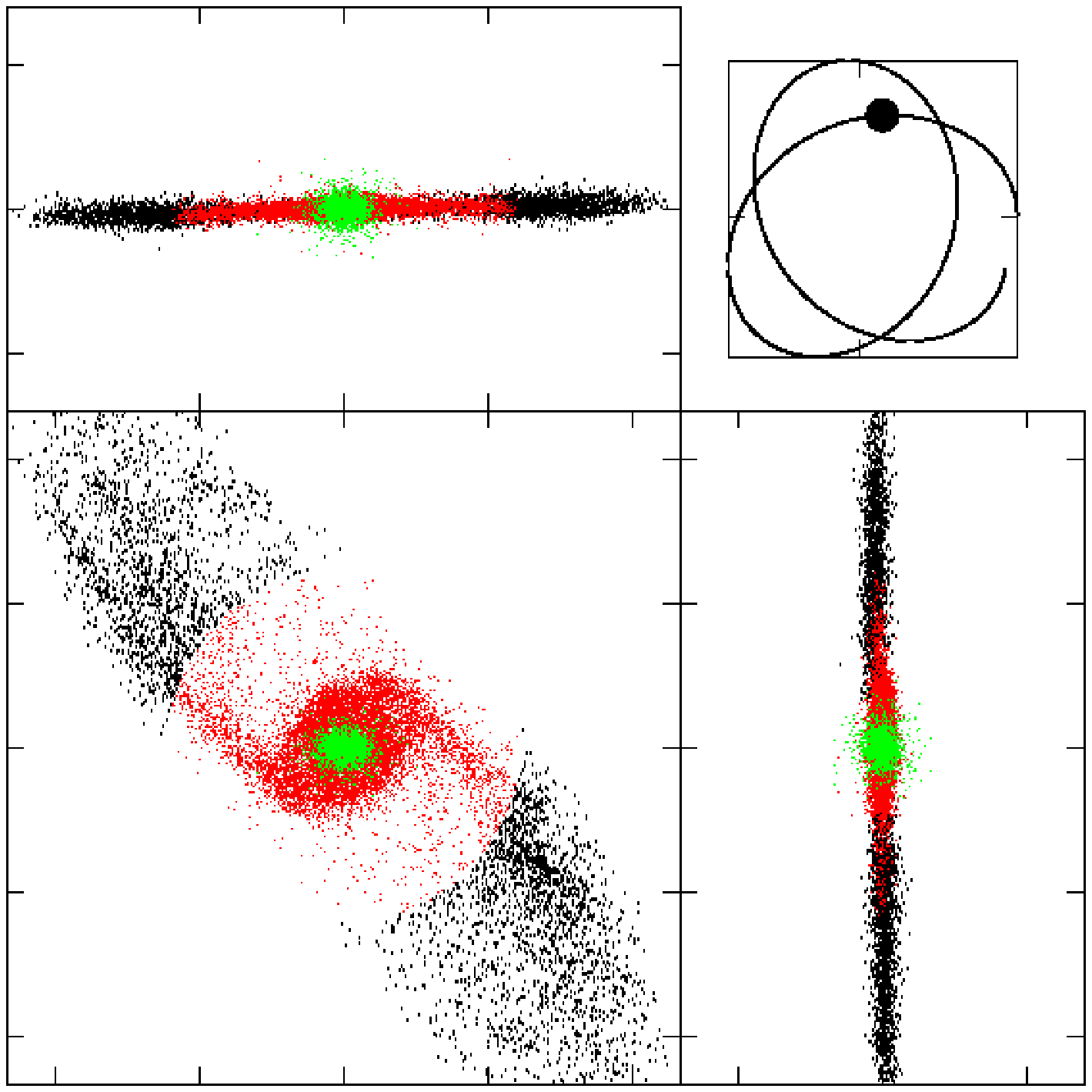}
           \includegraphics[angle=-90. ,width=0.3\textwidth]{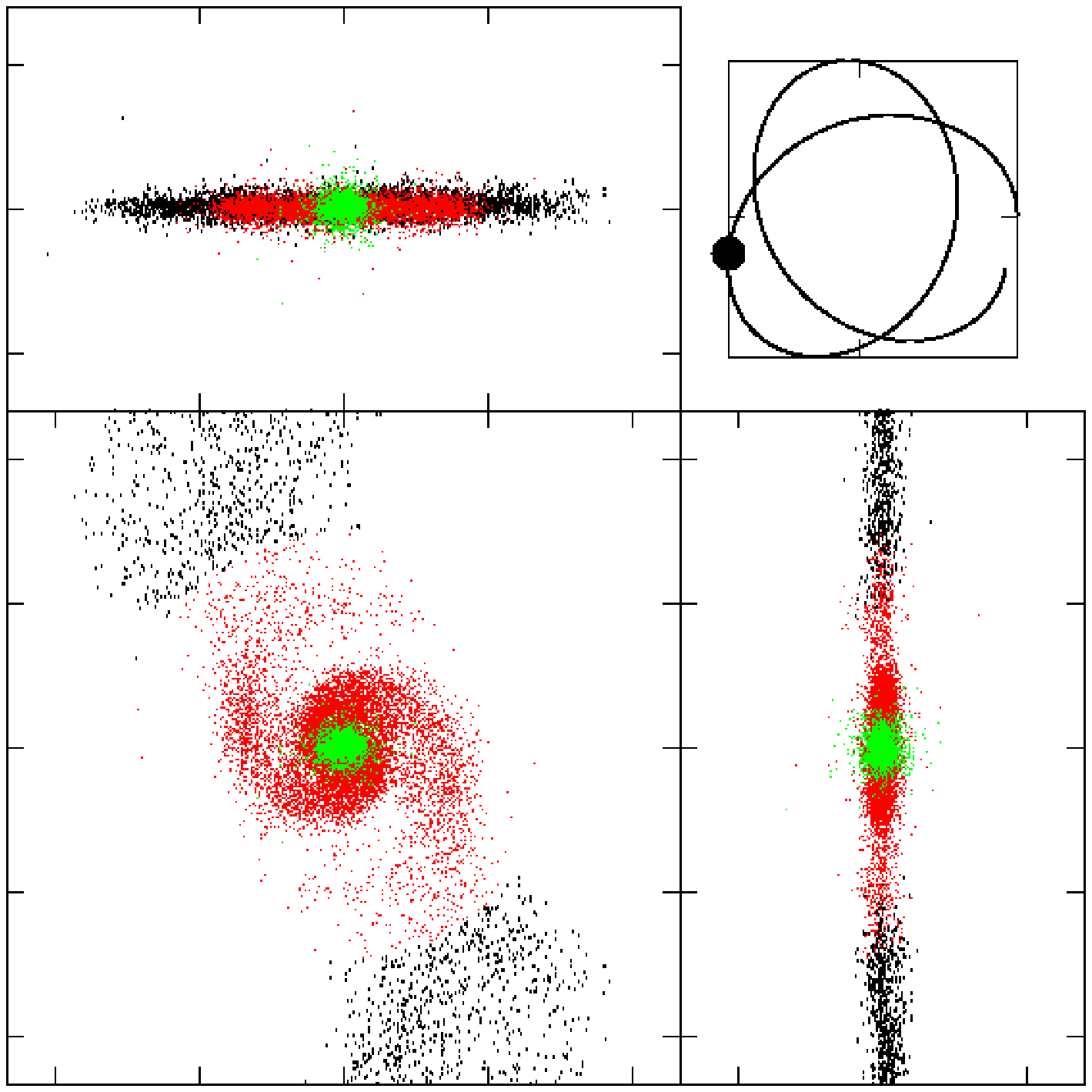}}     
        \subfigure{ 
           \includegraphics[angle=-90. ,width=0.363\textwidth]{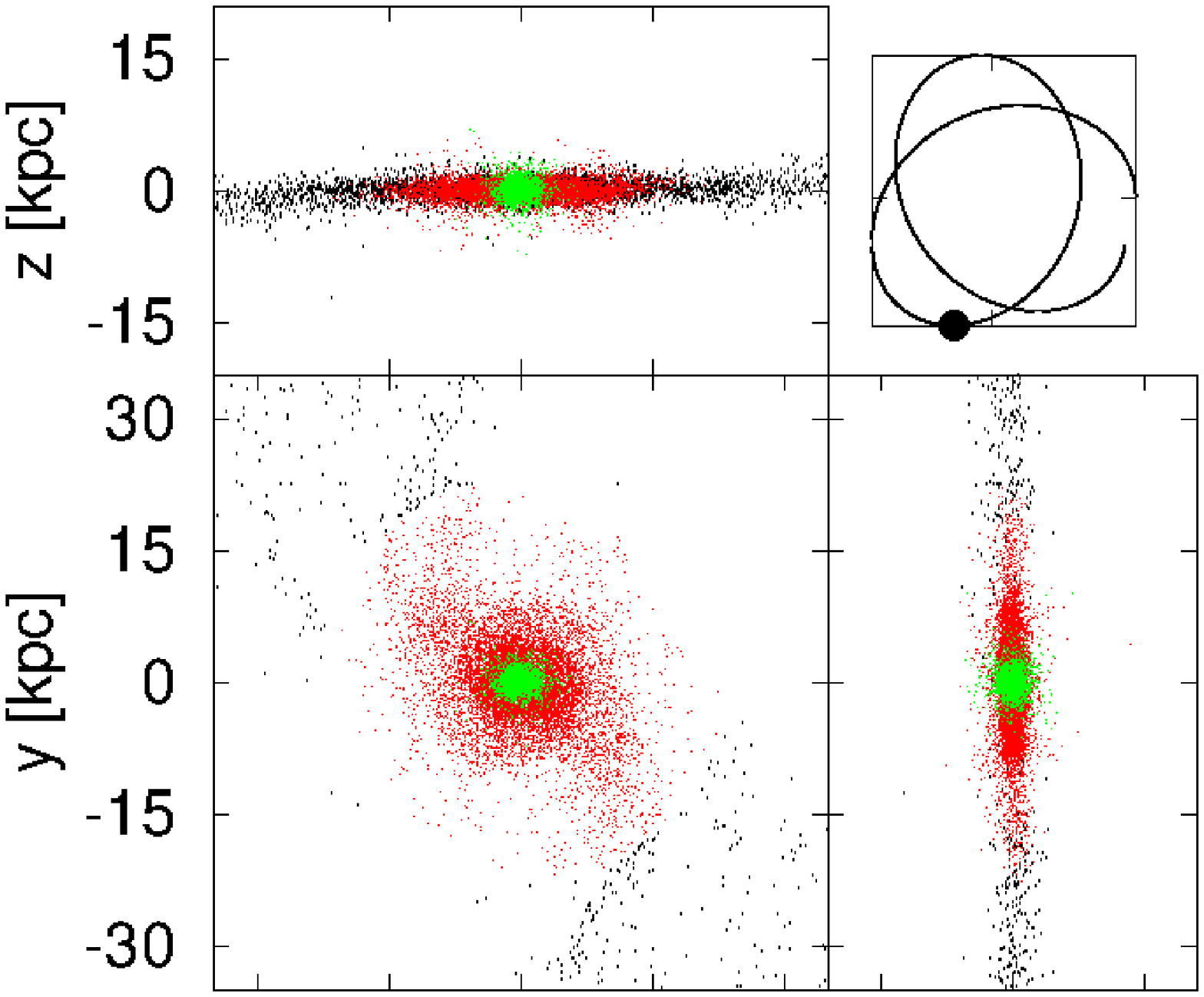}
           \includegraphics[angle=-90. ,width=0.3\textwidth]{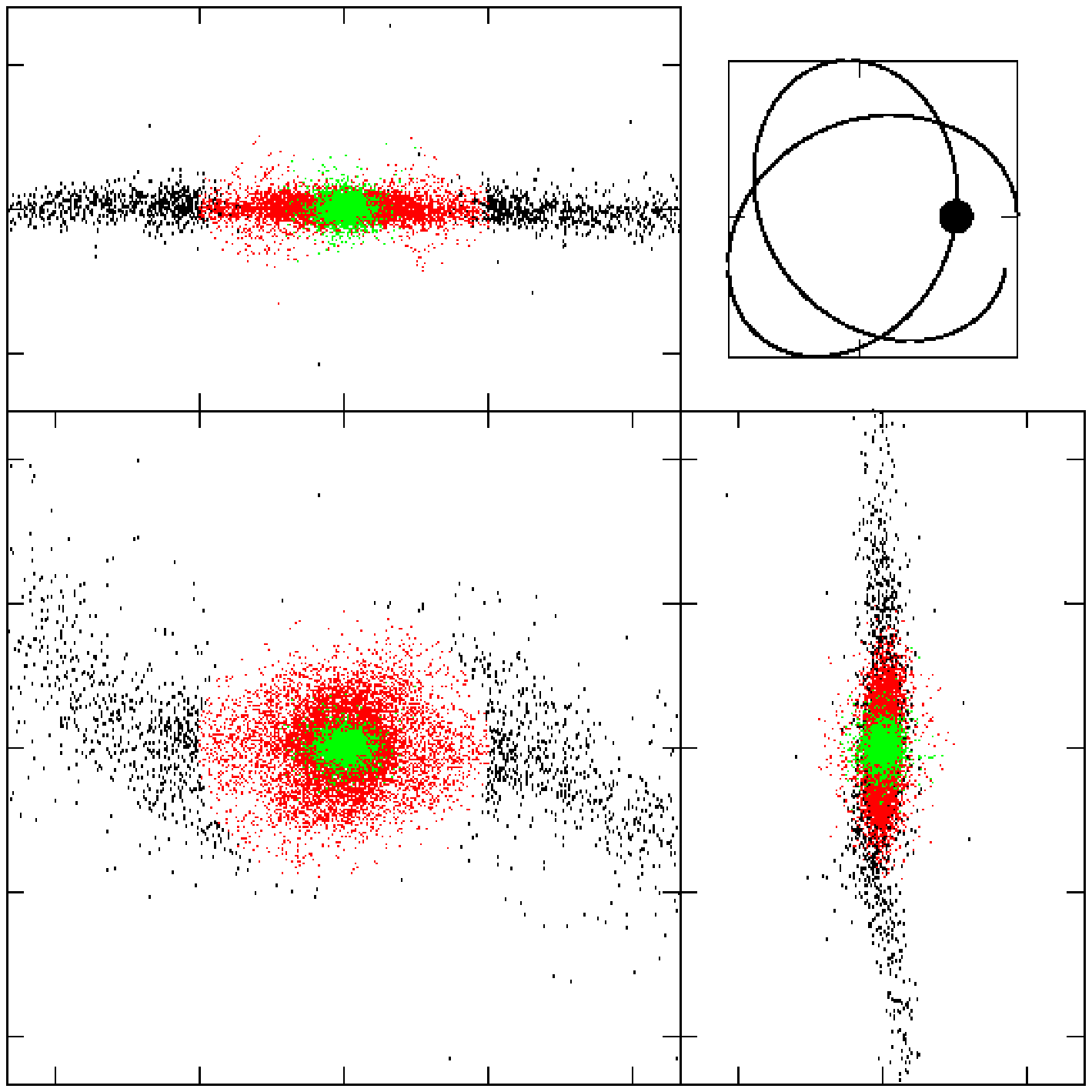}
           \includegraphics[angle=-90. ,width=0.3\textwidth]{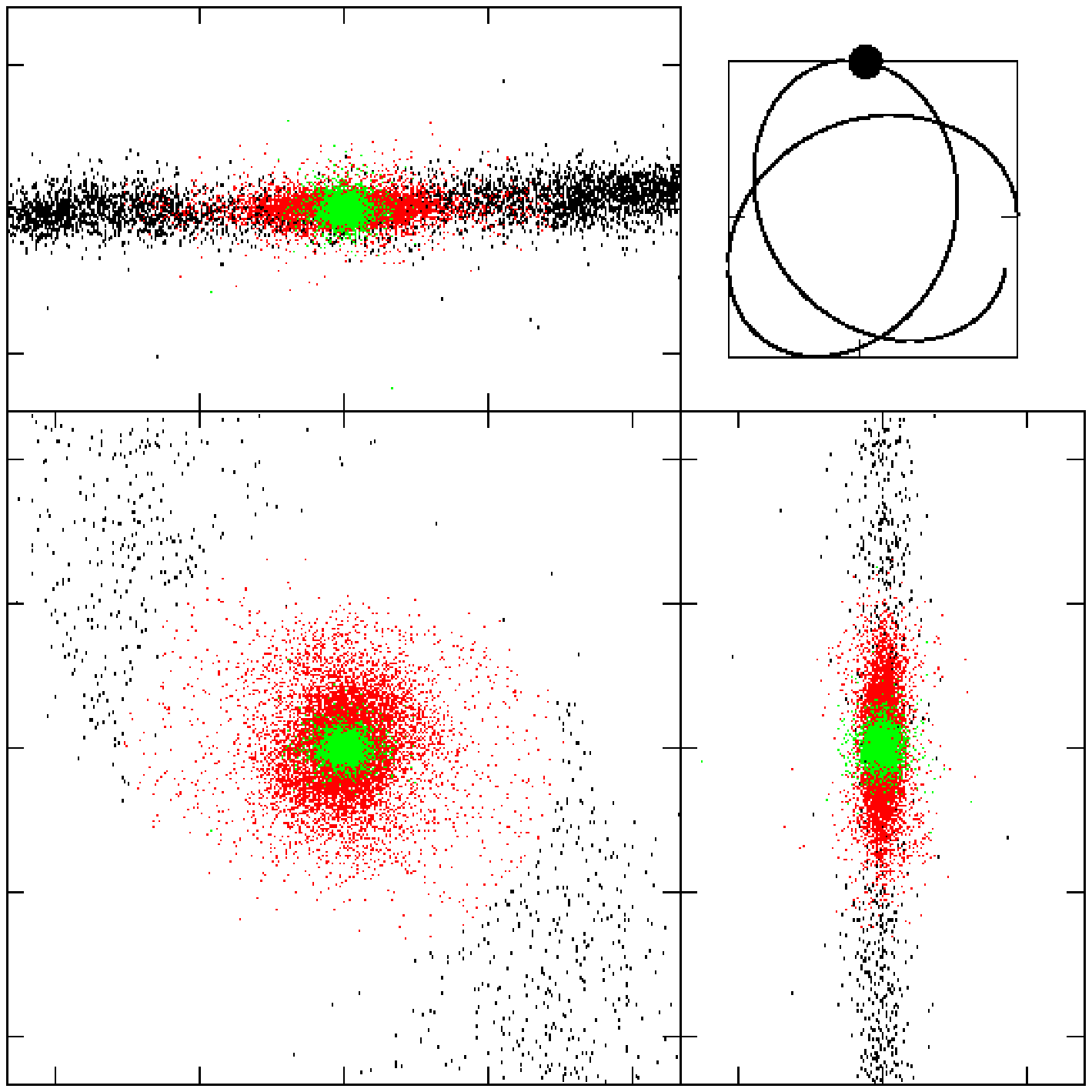}}                
   \subfigure{ 
           \includegraphics[angle=-90. ,width=0.363\textwidth]{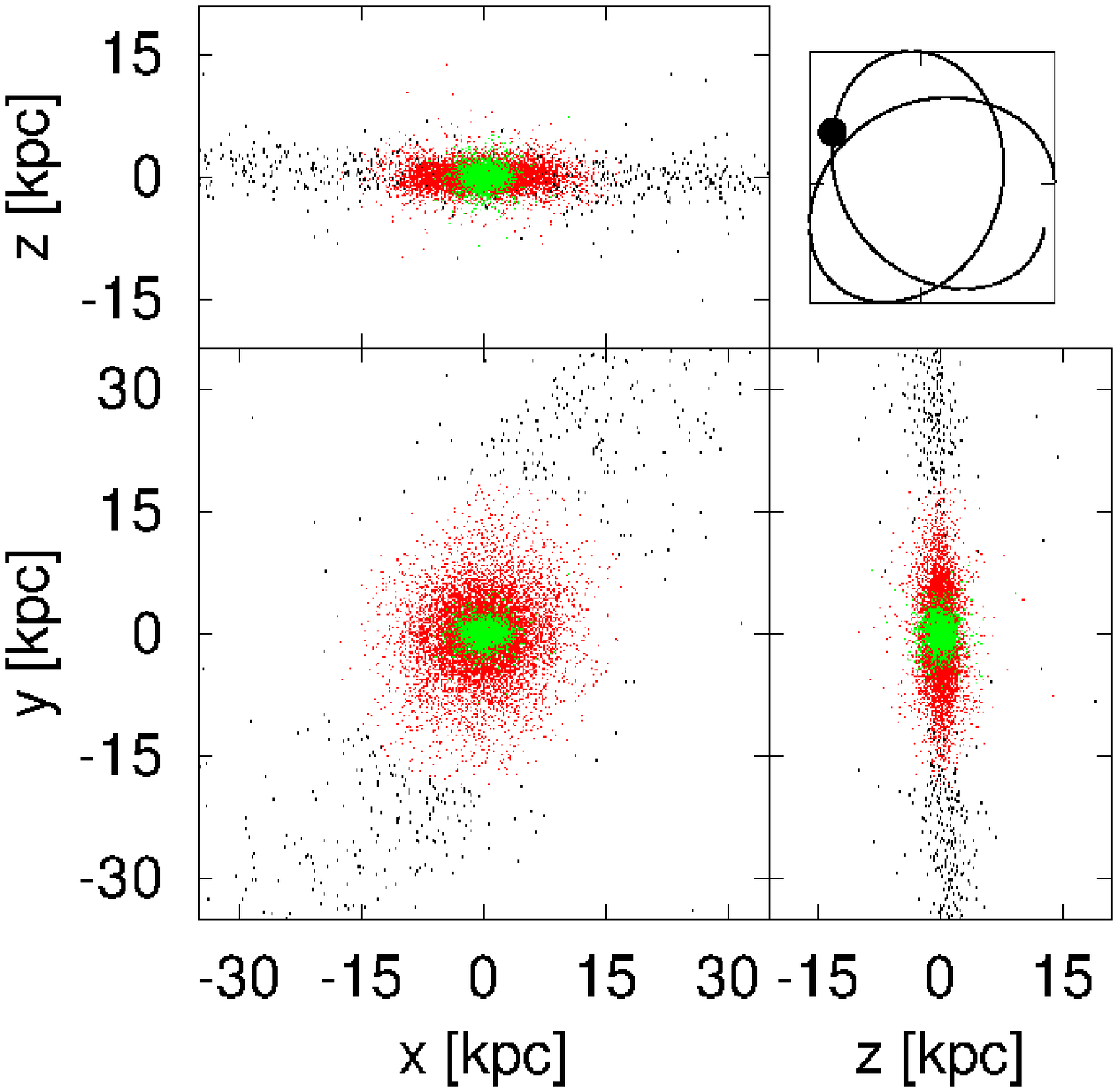}
           \includegraphics[angle=-90. ,width=0.3\textwidth]{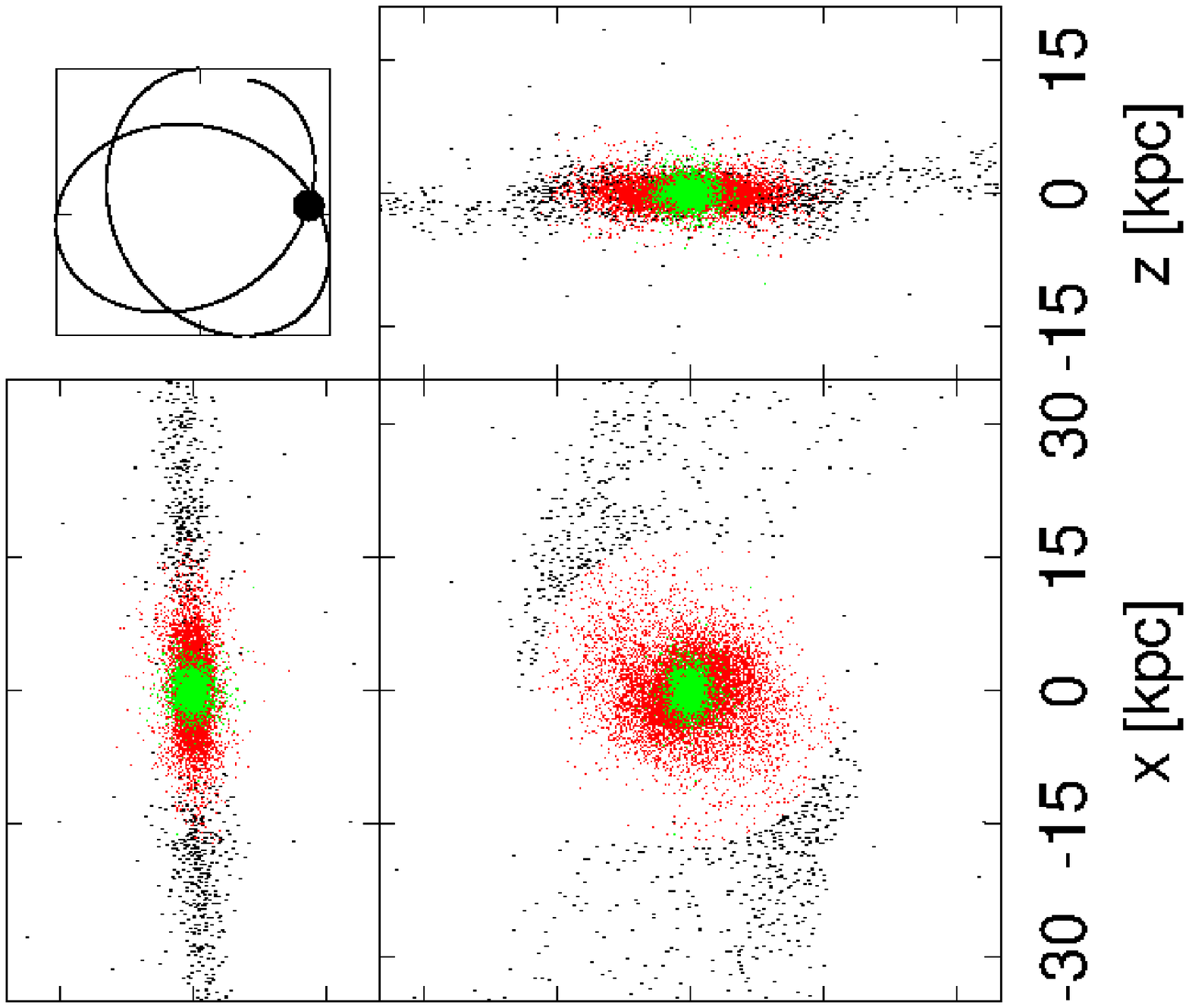}
           \includegraphics[angle=-90. ,width=0.3\textwidth]{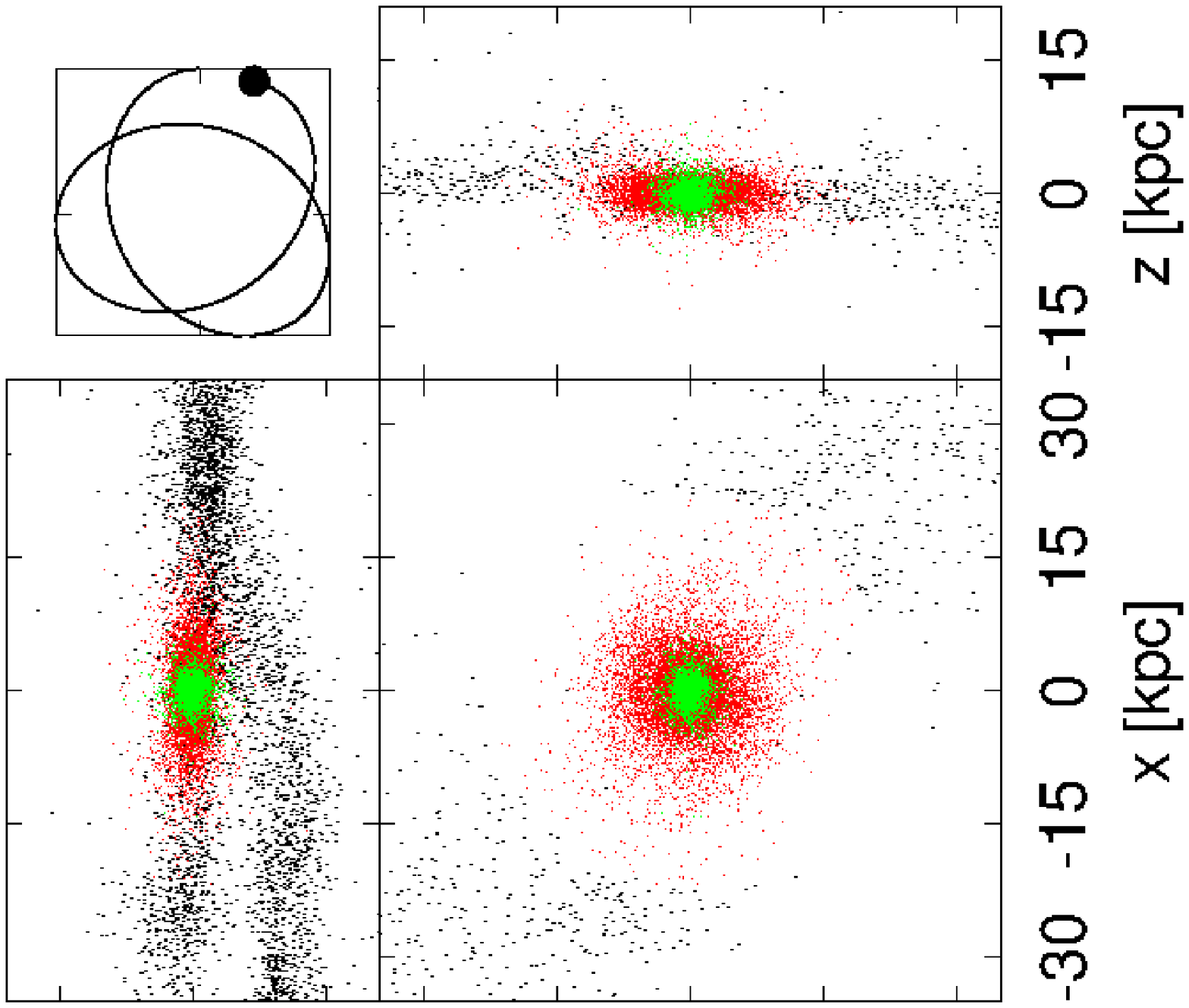}}     
    \caption{Snapshots from the first simulation run of the setup with the closest orbit, i.e.\ an apocentre of 0.5 Mpc and pericentre of 0.3 Mpc. The timestep between snapshots is 625 Myr; they start at the upper left at 0 Gyr and reach the lower right after 5 Gyr. Each snapshot image shows the galaxy face-on and from two perpendicular edge-on viewpoints. In the upper right corner of each image, the actual position on the orbit around the cluster centre is marked.
     The red dots are bound disk particles, the green dots are bound bulge particles, and the black dots are unbound stellar particles.}
    \label{SnapShots-1-1}
    \end{figure*}
    
   \begin{figure*}
   \subfigure{ 
           \includegraphics[angle=-90. ,width=0.363\textwidth]{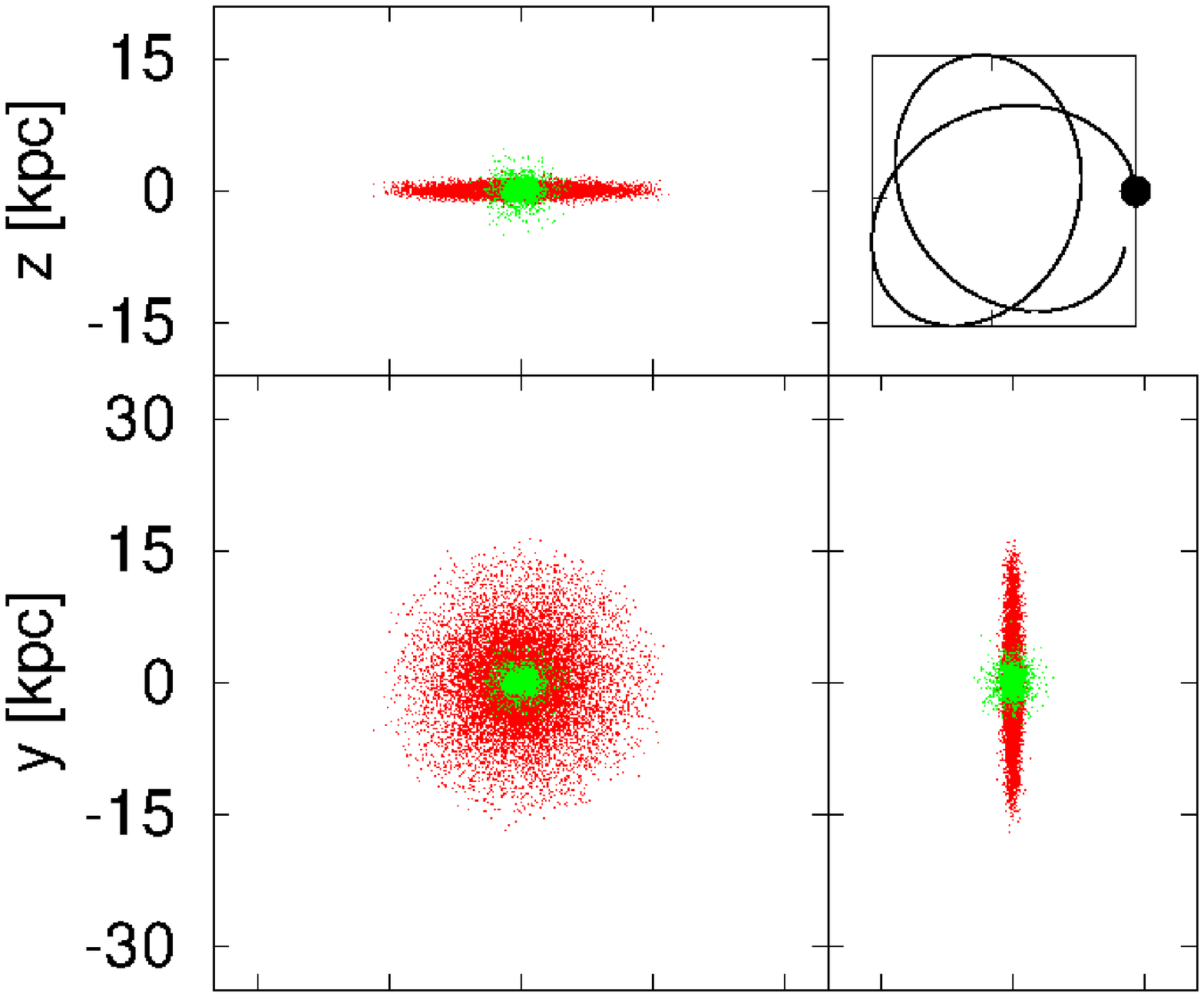}
           \includegraphics[angle=-90. ,width=0.3\textwidth]{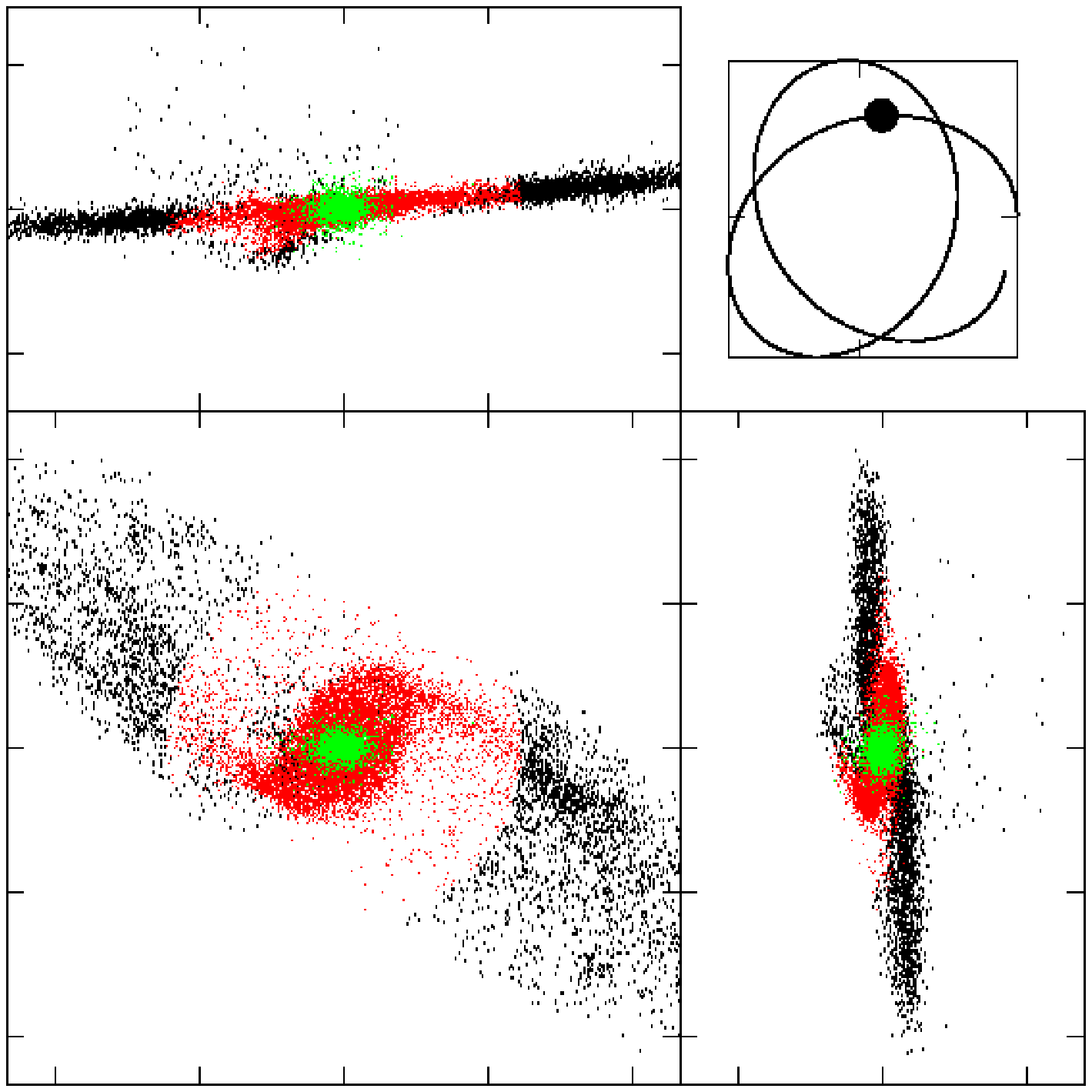}
           \includegraphics[angle=-90. ,width=0.3\textwidth]{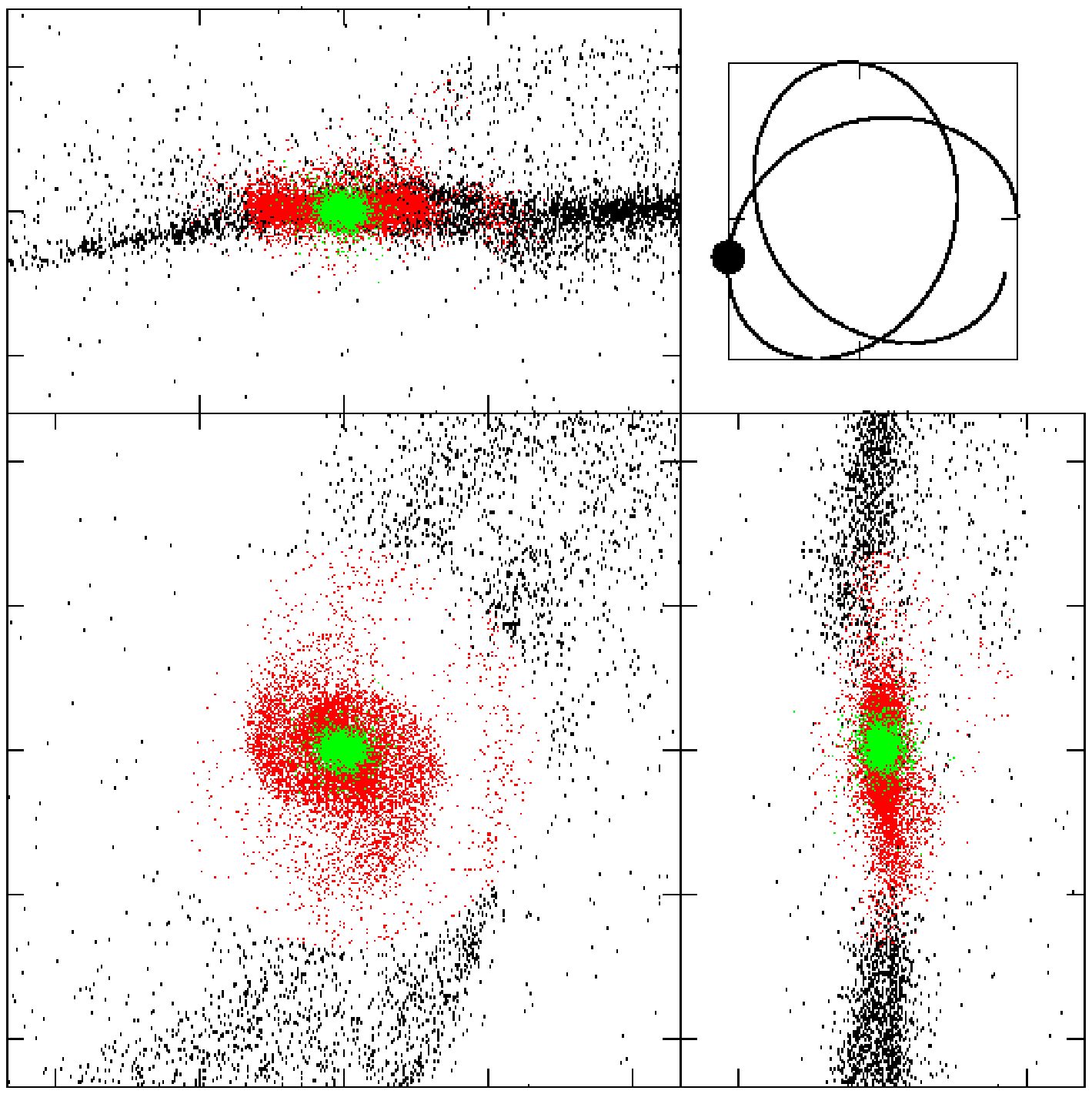}}     
        \subfigure{ 
           \includegraphics[angle=-90. ,width=0.363\textwidth]{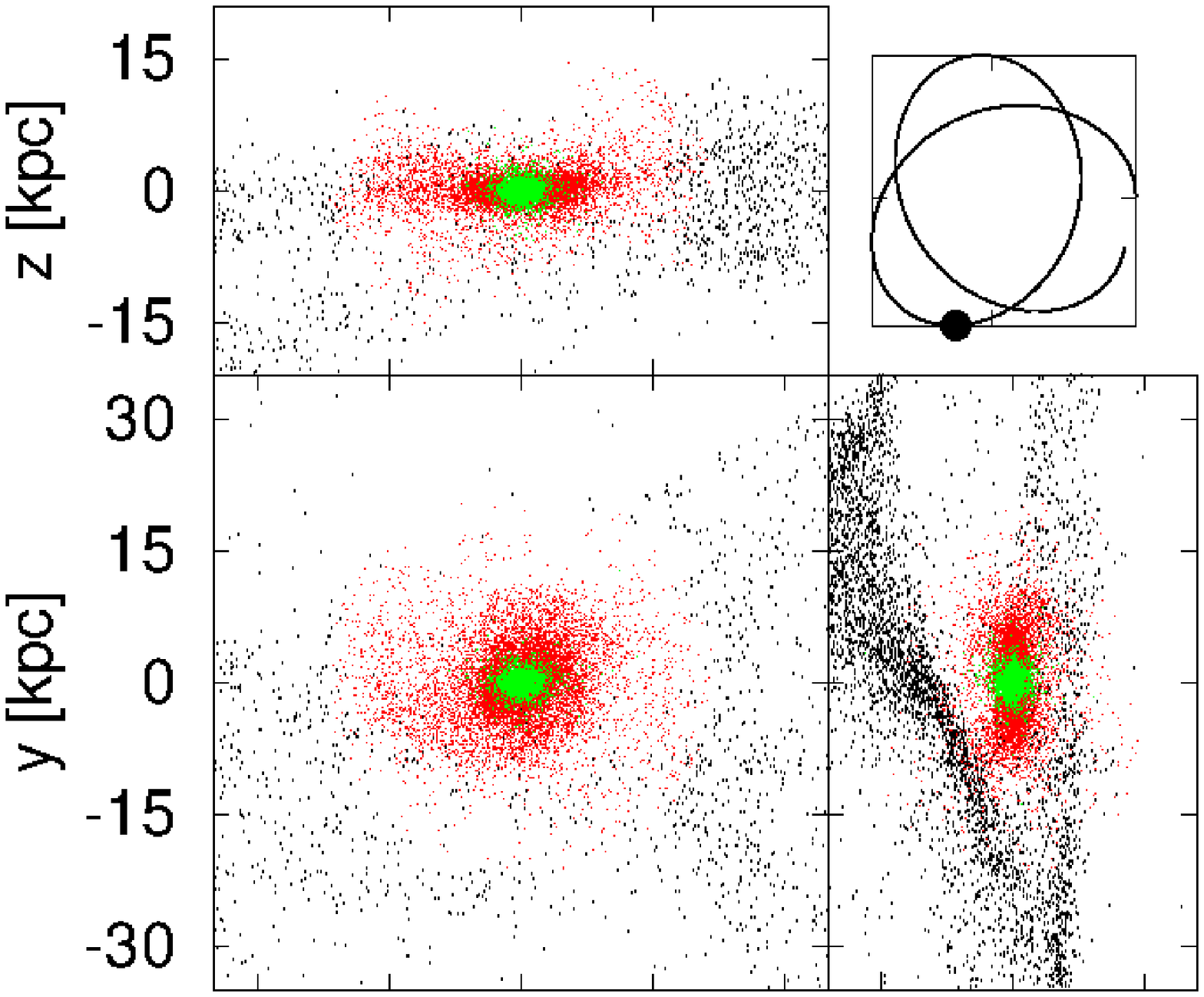}
           \includegraphics[angle=-90. ,width=0.3\textwidth]{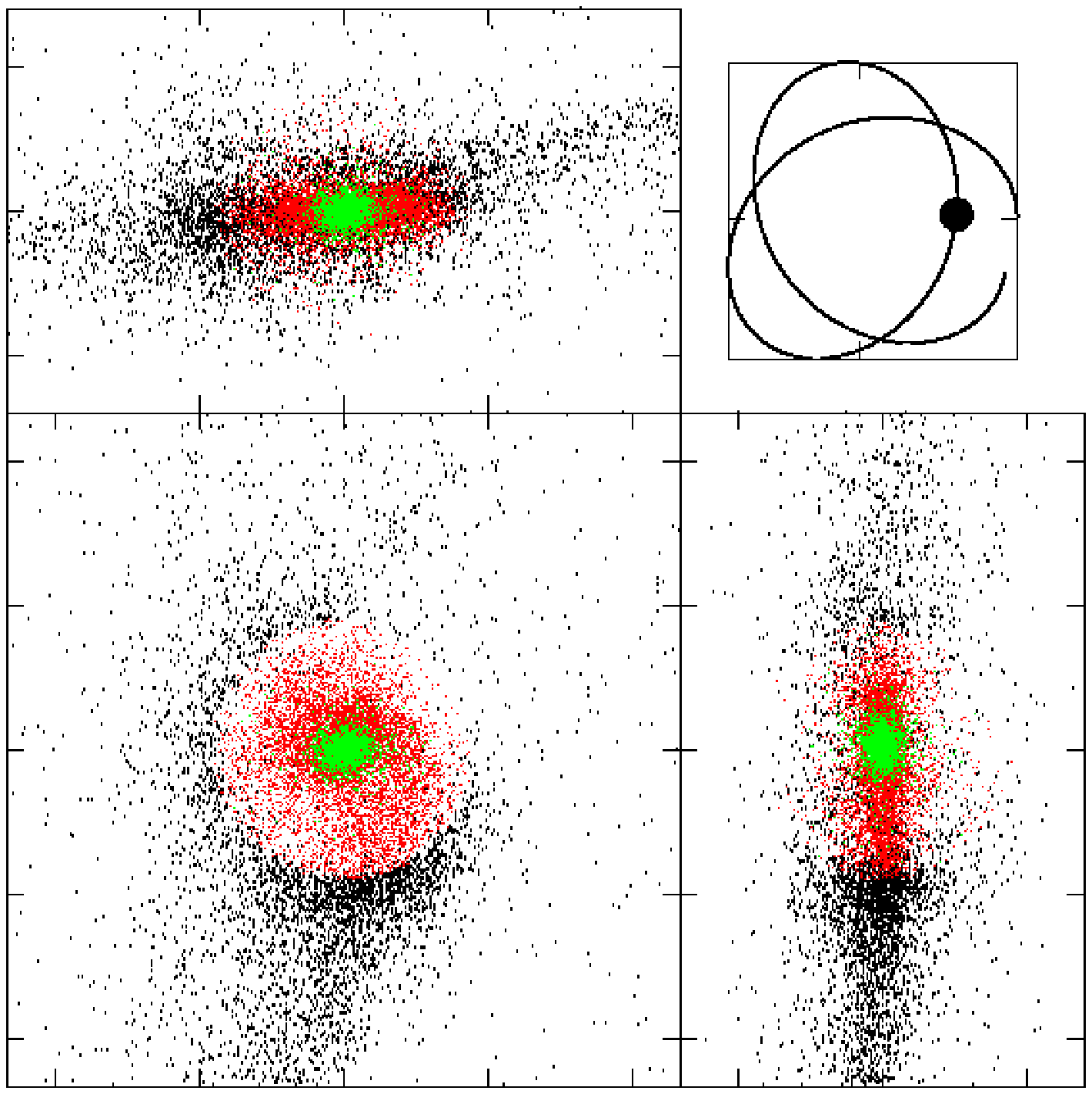}
           \includegraphics[angle=-90. ,width=0.3\textwidth]{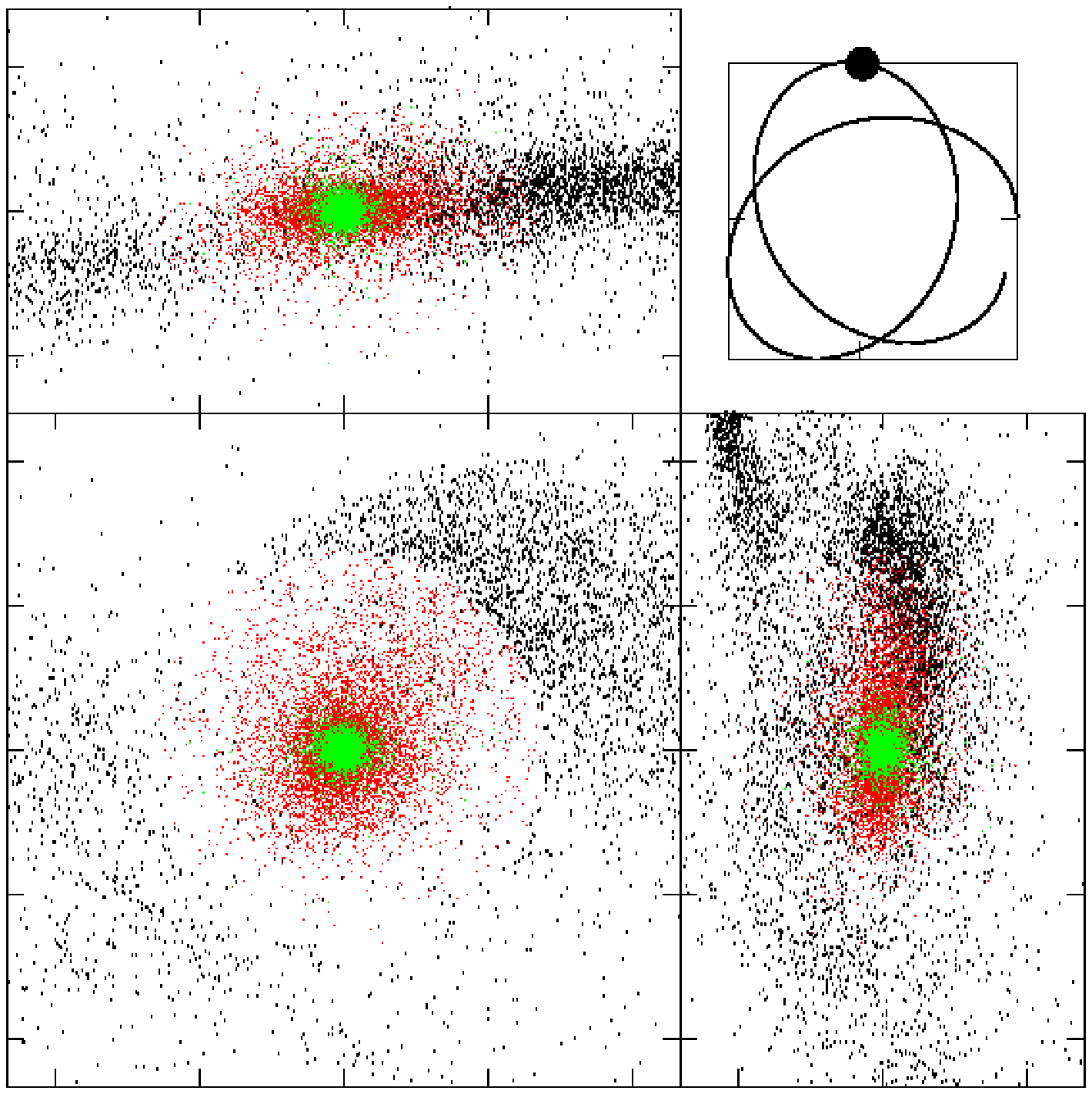}}                
   \subfigure{ 
           \includegraphics[angle=-90. ,width=0.363\textwidth]{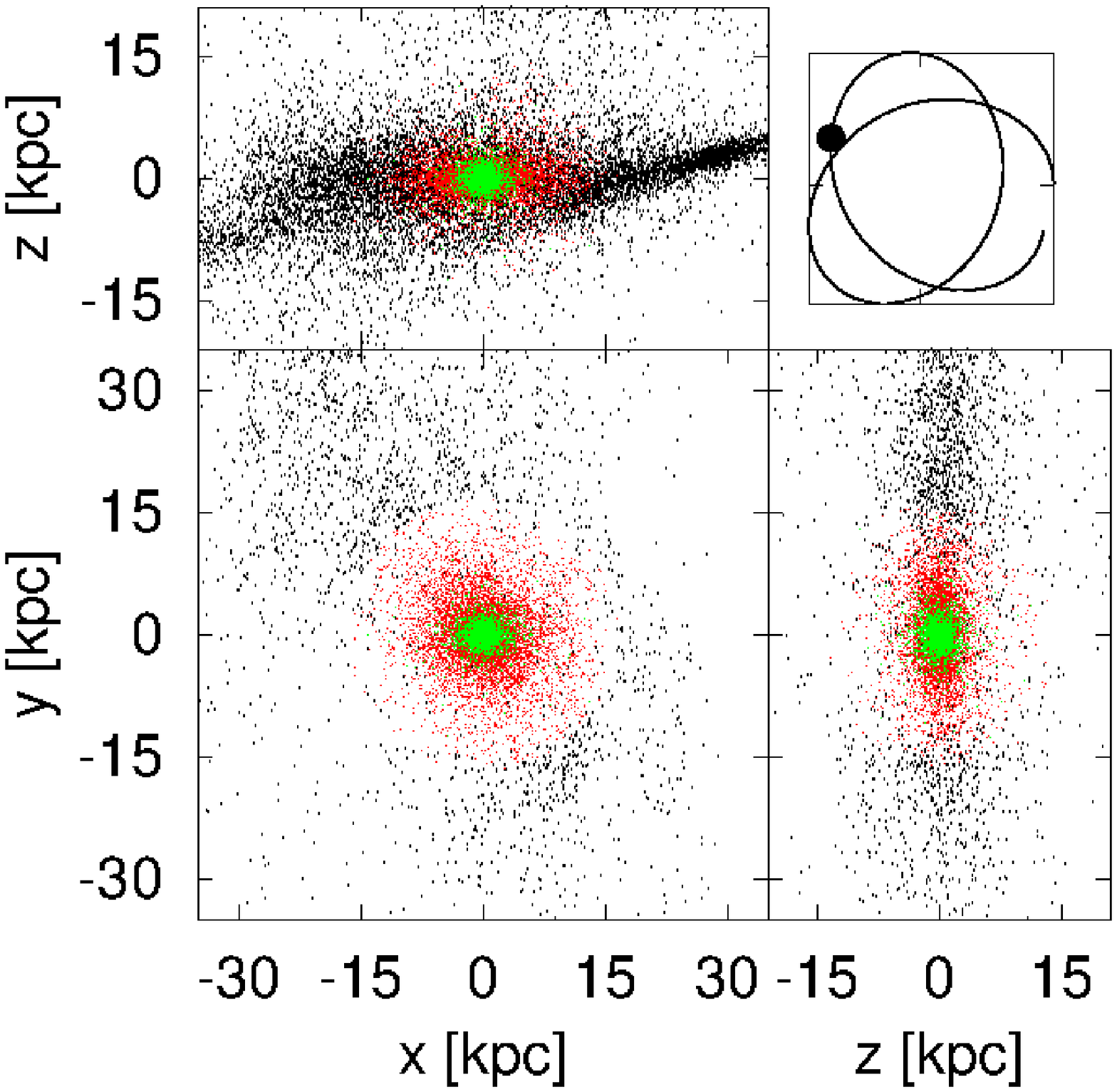}
           \includegraphics[angle=-90. ,width=0.3\textwidth]{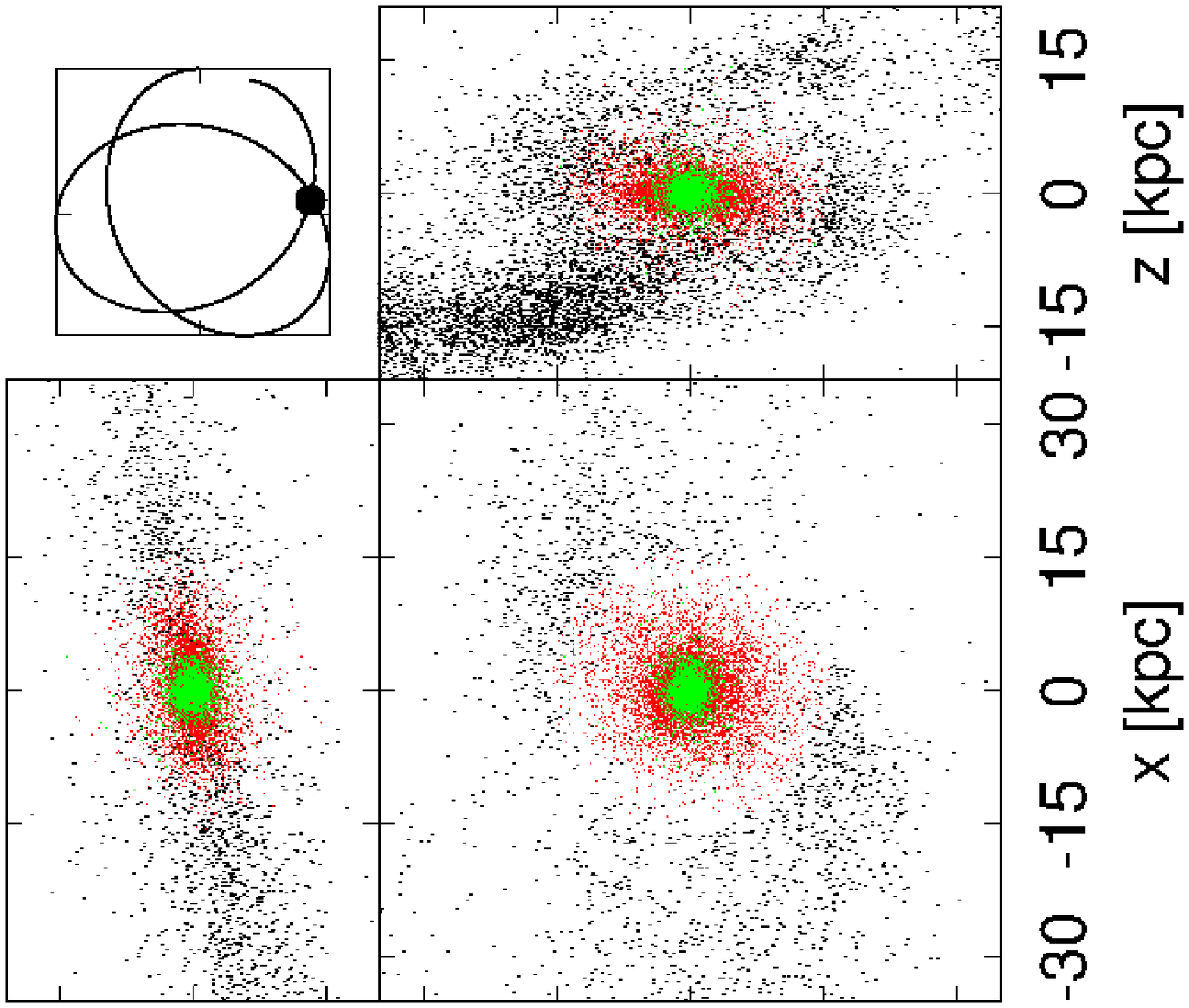}
           \includegraphics[angle=-90. ,width=0.3\textwidth]{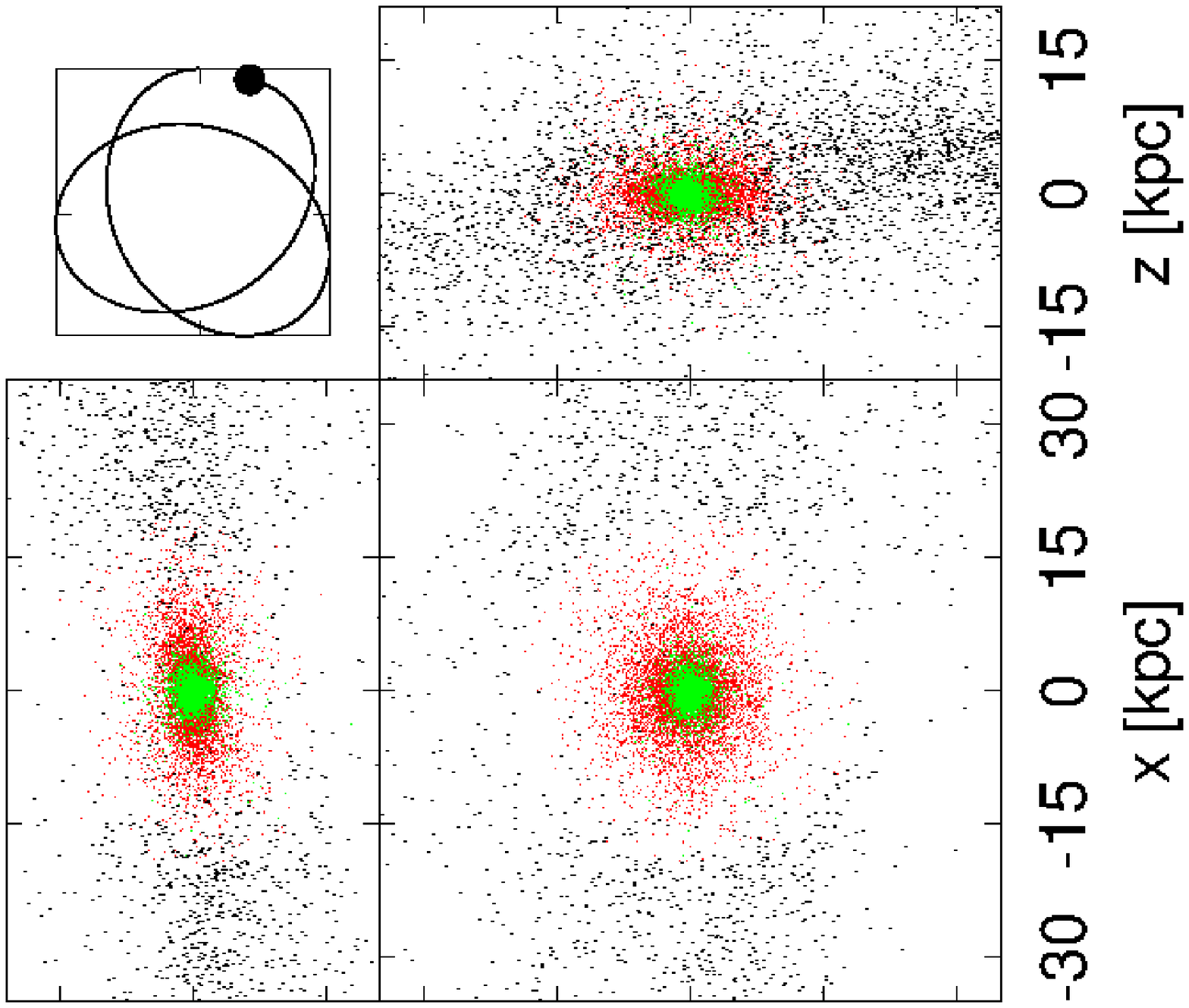}}     
    \caption{Like Fig.~\ref{SnapShots-1-1}, but for the second simulation run of the setup with the closest orbit.}
    \label{SnapShots-1-2}
    \end{figure*}

\begin{table*}
\caption{Simulation results}             
\label{table:3}      
\centering                          
\begin{tabular}{ c c c c | c c c c c c c c c c c}    
\hline\hline  
\multicolumn{4}{c|}{Setup} & \multicolumn{11}{c}{Galaxy properties} \\ 
\hline               
Apo. & Inc. & Model & Enc. & $  \langle R_{\rm T} \rangle / \langle R_{\rm 1/2} \rangle $ & $ \# enc$ & $  \# enc_{\Delta E > 10^{-2}}$ & Star & Tot & $R_{\rm 1/2}$ & $ <z>$ & $ c/a$ & $v_{\rm max}$ & $ \sigma$ & $v_{\rm max} / \sigma$ \\
\hline\hline
\multicolumn{9}{c}{Properties of the initial models} \\
\hline
- & - & large & -  & -  & -  & - & 100 & 100 & 5.00 & 0.37 & 0.08 & 65 & 25 & 2.60 \\ 
- & - & small & -  & -  & -  & - & 100 & 100 & 2.00  & 0.16 & 0.09 & 89 & 39 & 2.28 \\
\hline\hline
\multicolumn{9}{c}{Properties of the galaxy remnants} \\ 
\hline
0.5 & $0^{\circ}$ & large & y & 3.7 & 15 & 0 & 77 & 46 & 3.40 & 0.60 & 0.16 & 48 & 39 & 1.22 \\
0.5 & $0^{\circ}$ & large & y & 3.7 & 20 & 1 & 66 & 32 & 3.10 & 1.05 & 0.32 & 30 & 36 & 0.84 \\
0.5 & $0^{\circ}$ & large & y & 3.8 & 18 & 1 & 55 & 30 & 2.50 & 1.12 & 0.43 & 29 & 43 & 0.68 \\
0.5 & $0^{\circ}$ & large & n & 3.7 & 0  & 0 & 75 & 46 & 3.50 & 0.59 & 0.17 & 48 & 37 & 1.28 \\

1.0 & $0^{\circ}$ & large & y & 5.3 & 14 & 0 & 92 & 39 & 3.90 & 0.77 & 0.15 & 46 & 41 & 1.12 \\ 
1.0 & $0^{\circ}$ & large & y & 5.5 & 14 & 0 & 89 & 37 & 3.60 & 0.85 & 0.19 & 42 & 39 & 1.07 \\
1.0 & $0^{\circ}$ & large & y & 5.3 & 13 & 1 & 87 & 35 & 3.90 & 1.41 & 0.21 & 40 & 36 & 1.12 \\ 
1.0 & $0^{\circ}$ & large & n & 5.3 & 0 & 0 & 92 & 40 & 3.90 & 0.67 & 0.11 & 47 & 39 & 1.20 \\ 

1.5 & $0^{\circ}$ & large & y & 7.8 & 11 & 0 & 96 & 40 & 4.20 & 0.59 & 0.10 & 52 & 36 & 1.45 \\ 
1.5 & $0^{\circ}$ & large & y & 8.0 &  9 & 0 & 94 & 37 & 3.90 & 0.61 & 0.10 & 52 & 37 & 1.39 \\
1.5 & $0^{\circ}$ & large & y & 7.8 &  5 & 0 & 96 & 38 & 4.10 & 0.58 & 0.10 & 53 & 36 & 1.48 \\ 
1.5 & $0^{\circ}$ & large & n & 7.8 &  0 & 0 & 96 & 40 & 4.10 & 0.53 & 0.08 & 53 & 32 & 1.64 \\ 

0.5 & $0^{\circ}$ & small & y & 8.4 & 42 & 0 & 100 & 94 & 2.00 & 0.17 & 0.09 & 87 & 41 & 2.00 \\
0.5 & $0^{\circ}$ & small & y & 8.4 & 44 & 0 &  98 & 77 & 2.00 & 0.43 & 0.19 & 81 & 41 & 1.99 \\
0.5 & $0^{\circ}$ & small & y & 8.4 & 45 & 0 & 100 & 94 & 2.00 & 0.17 & 0.09 & 86 & 41 & 2.12 \\
0.5 & $0^{\circ}$ & small & n & 8.4 &  0 & 0 & 100 & 94 & 2.00 & 0.16 & 0.08 & 87 & 41 & 2.14 \\

0.5 & $45^{\circ}$ & large & y & 3.6 & 13 & 0 & 80 & 46 & 3.60 & 1.11 & 0.41 & 50 & 37 & 1.33 \\ 
0.5 & $45^{\circ}$ & large & y & 3.4 & 18 & 1 & 28 & 14 & 2.80 & 1.37 & 0.49 & 22 & 29 & 0.76 \\
0.5 & $45^{\circ}$ & large & y & 3.8 & 21 & 0 & 72 & 41 & 3.20 & 1.43 & 0.54 & 43 & 36 & 1.20 \\ 
0.5 & $45^{\circ}$ & large & n & 3.7 & 0  & 0 & 80 & 46 & 3.50 & 1.11 & 0.42 & 50 & 37 & 1.33 \\ 

0.5 & $90^{\circ}$ & large & y & 3.4 & 19 & 0 & 88 & 51 & 4.00 & 0.81 & 0.29 & 59 & 31 & 1.93 \\  
0.5 & $90^{\circ}$ & large & y & 3.4 & 21 & 1 & 77 & 43 & 3.60 & 1.47 & 0.34 & 42 & 36 & 1.17 \\ 
0.5 & $90^{\circ}$ & large & y & 3.4 & 20 & 2 & 62 & 32 & 3.10 & 1.86 & 0.55 & 34 & 37 & 0.91 \\ 
0.5 & $90^{\circ}$ & large & n & 3.4 & 0  & 0 & 89 & 49 & 4.00 & 0.89 & 0.33 & 60 & 31 & 1.96 \\ 

\hline                                   
\end{tabular}
\tablefoot{(1) apocentre of the orbit in Mpc, the pericentre was always fixed to 0.3Mpc;  (2) inclination of the orbital plane and the disk; (3) used galaxy model;  (4) enc. indicates if a simulation is using encounters with cluster members or not; (5) ratio of the averaged tidal radius and the half-mass radius; (6) and (7) are the number of galaxy-galaxy encounters in this simulation and the number of galaxy-galaxy encounters with an approximated energy transfer greater than $10^{-2}$; (8) and (9) fraction of the initially bound stellar and total mass; (10) half-mass radius of the galaxy in kpc; (11) mean height of stars above the disk in kpc; (12) axis ratio of the stellar component; (13) maximum rotation velocity in km/s; (14) velocity dispersion in km/s; (15) ratio of peak velocity and mean velocity dispersion. All the values have already been corrected for relaxation as explained in section 2.3}
\end{table*}

\subsection{Effect of orbital parameters and encounters}

As expected, the orbital parameters were essential for the evolution of a galaxy inside of a galaxy cluster environment. The evolution of the bound stellar mass fractions in the simulations with different apocentres (0.5Mpc, 1.0Mpc, and 1.5Mpc) is shown in fig.~\ref{MstarOrb}. On the farthest orbit, the stellar mass loss was between $4\%$  and  $ 6\%$, whereas on the innermost orbit, the stellar mass loss was $24\%$ to $50\%$. On the intermediate orbit, it reached $8\%$ to $13\%$. The scatter of the mass loss in the different setups is a result of the stochastical nature of the harassment scenario.

Figures~\ref{SnapShots-1-1} and \ref{SnapShots-1-2} show the evolution of the galaxies in the first and second runs of the setup with the closest orbit by a series of snapshots with intervals of 625 Myr. In the first run of this setup, no encounter with another cluster member resulted in additional mass loss. As opposed to that, in the second run at a very early stage of the simulation, an encounter pulled nearly $5\%$ of all stellar particles out of the disk and disturbed the system strongly. 
This resulted in a different mass loss history and different shape of the remnant.
For a better understanding of the mass loss histories and evolution of the galaxies, it is reasonable to separate the effects of the global tidal field and close encounters, as we do in the following.

\subsubsection{Mass loss on different orbits without encounters}

Using the simulations without encounters we could follow the evolution of the infalling galaxy introduced only by the global tidal field of the cluster. This evolution is a kind of minimal effect that each galaxy experiences if it falls on a given orbit into a cluster. That effect could be amplified by encounters, but not weakened. 

\begin{figure}
   \centering
   \includegraphics[angle=-90. ,width=0.5\textwidth]{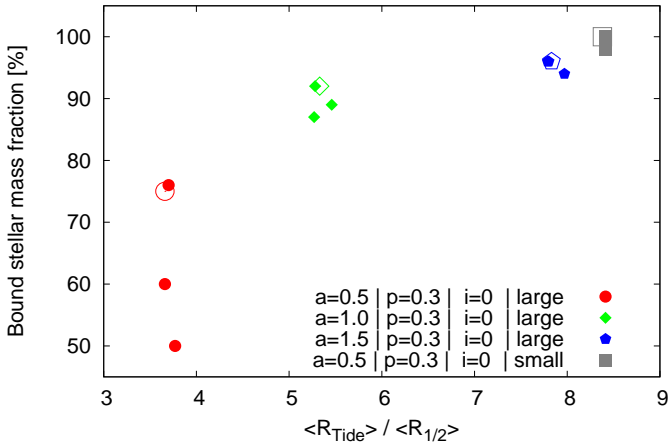}
   \caption{Bound stellar mass fraction versus the averaged tidal radius in units of the averaged half-mass radius of the model galaxy. Open symbols represent simulations without encounters with other cluster members.}
   \label{M_Star-R_tide_onlyOrbite2}
   \end{figure}

The stellar mass loss by the global tidal field (dotted lines in Fig.~\ref{MstarOrb}) started when the galaxy passed the pericentre the first time. At the pericentre the tidal forces reached their maximum strength and therefore the tidal radius reached a local minimum. The local minimum of the tidal radius is equivalent to a local minimum in the bound mass fraction.  The small tidal radius at the pericentre enabled the growth of tidal tails. Along these tidal tails, stars also left the galaxy more than one Gyr after the galaxy passed the pericentre. The smaller the apocentre of an orbit, the more frequently the galaxy passes the pericentre and the smaller the tidal radius averaged over the simulation time. The tidal force becomes more efficient in removing material the smaller the averaged tidal radius gets relative to the averaged half-mass radius of the stellar component of the galaxy. Figure~\ref{M_Star-R_tide_onlyOrbite2} shows that for the galaxies without an inclination, the ratio of the averaged tidal radius along the orbit and the averaged half-mass radius of the stellar component is an adequate way to quantify the efficiency of tidal forces. The mass loss caused by the persistent global tidal forces of the cluster's background mass represents the minimum mass loss by the harassment scenario. \\
 A huge part of the dark matter halo was also stripped at the first pericentre passage. For the large model, the loss of initially bound total mass was close to or more than $50\%$, for the small model this loss was in most cases only $6\%$.

\subsubsection{Mass loss on different orbits with encounters}

\begin{figure}
   \centering
   \includegraphics[angle=-90. ,width=0.5\textwidth]{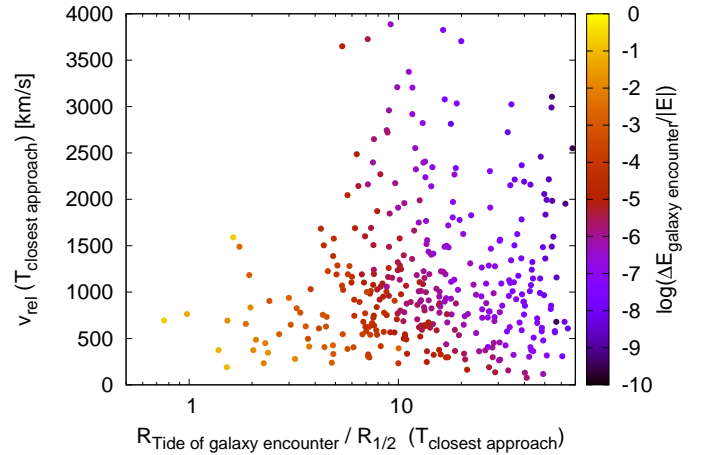}
   \caption{Relative velocity of the flying-by cluster members versus the tidal radius with respect to the flying-by cluster members in units of the half-mass radius of the model galaxy at the moment of closest approach. The colour scale denotes the approximate energy that the flying-by cluster members inject into the stellar component of the galaxy, in units of the total energy of the stellar component.} 
   \label{Enc_Rtide_vs_V_all}
   \end{figure}

Within one setup, the runs show a large scatter in their stellar mass loss. This scatter results from different encounters with cluster members. Not every encounter has the same impact, it depends on the minimal distance, the mass ratio and the relative velocity of both galaxies. 
To describe the impact of encounters in a similar way to how we described the impact of the global tidal field before, we calculated the tidal radius of the galaxy with respect to the flying-by objects at the moment of closest approach. This tidal radius is expressed in units of the half-mass radius of the stellar component of the galaxy model. This quantity shows whether a flying-by cluster member could have any influence on the galaxy. However, one also needs to take the relative velocity of both galaxies into account, because the slower a galaxy is as it fies by, the higher the energy and angular momentum transfer from this object.
Figure~\ref{Enc_Rtide_vs_V_all} shows the distribution of the above properties for all flying-by galaxies in our simulations. The symbol colour indicates the approximate energy that the flying-by cluster members inject into the stellar component of the galaxy, in units of the total energy of the stellar component. To calculate this deposited energy, we used the approximation of \cite{1958ApJ...127...17S} for impulsive high-speed encounters. This is a simplification of the real galaxy-galaxy encounter, but it is accurate enough to give an impression of the strength of the encounter.
The figure illustrates that only a small number of all encounters with other cluster members fulfilled both criteria: to have a strong influence and be slow in relative velocity. Consequently, only a few encounters result in significant energy injection and additional mass loss. 
Therefore several encounters with cluster members are needed to have a strong impact on the infalling galaxy. This small fraction of particularly strong encounters with cluster members is also why the mass loss curves of several simulations are close to the mass loss curves of simulations without encounters. The different evolution of a galaxy with and without strong encounter is illustrated in Figs. 3 and 4 for the same simulation snapshots. The total number of encounters and the number of strong encounters (with $\Delta E \geq 10^{-2}$) in each simulation run is given in columns (6) and (7) of Table~\ref{table:3}. The total number of encounters in simulations with the small-galaxy model is two times higher than for simulations with the large model on the same orbit, because for the encounter probability, we take cluster members with masses above $0.1 M_{\rm gal}$ into account($M_{\rm gal}$ is the initial mass of the galaxy model), therefore more galaxy encounters occur to the smaller and lighter model. But it is also harder to affect the small model by an encounter. It has to come closer to the galaxy to strip the disk stars of the smaller model away, and the probability of such close encounters declines quadratically with the impact parameter.

The scatter of mass loss within a given setup declines with the increasing apocentre of the orbit. The reason is that, in setups with larger apocentre, the galaxies have spent a smaller amount of time in regions of the cluster with a high galaxy density, and therefore fewer encounters occurred.

\subsubsection{Transformation of shapes and kinematics}

\begin{figure}
   \centering
   \includegraphics[angle=-90. ,width=0.5\textwidth]{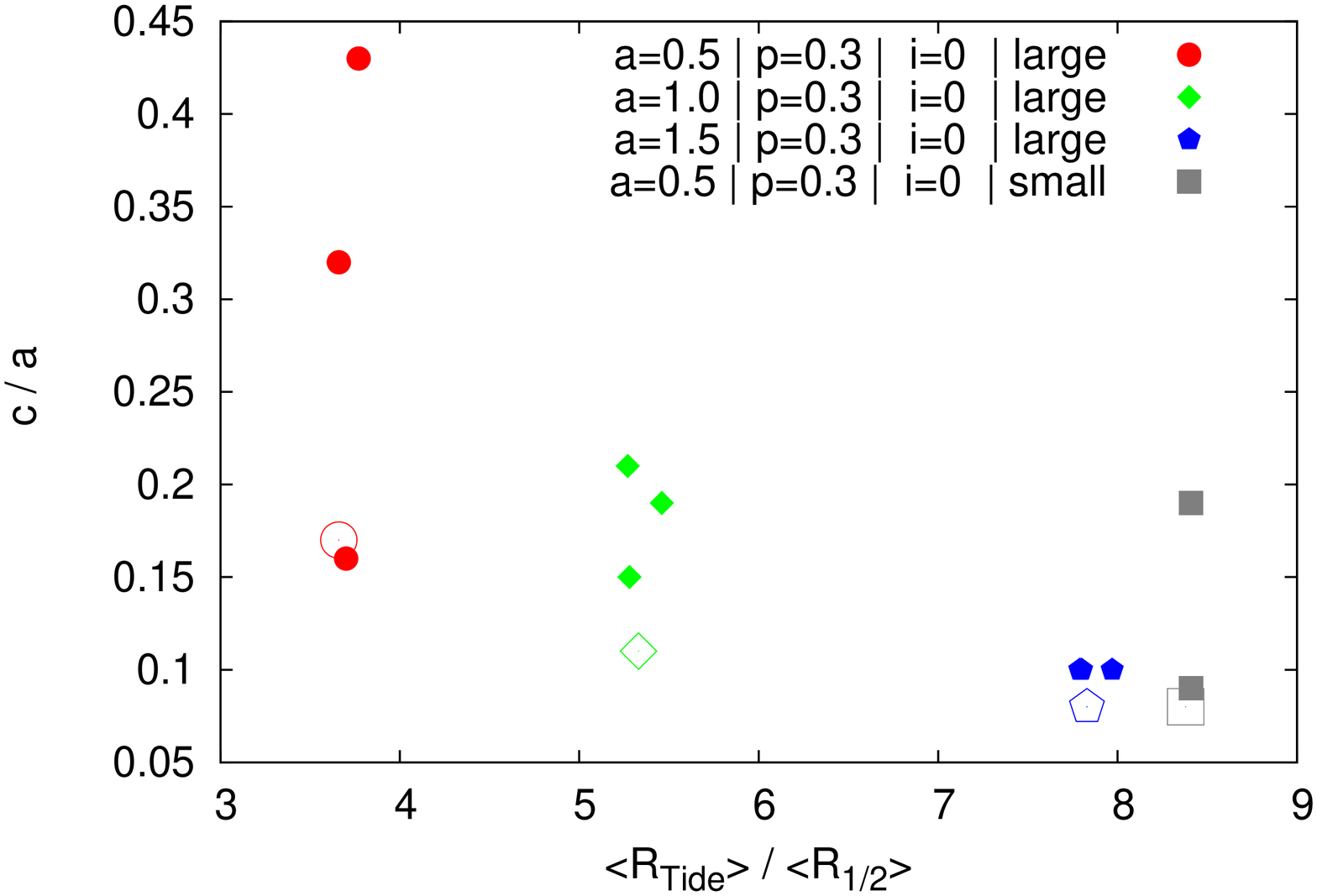}  
   \includegraphics[angle=-90. ,width=0.5\textwidth]{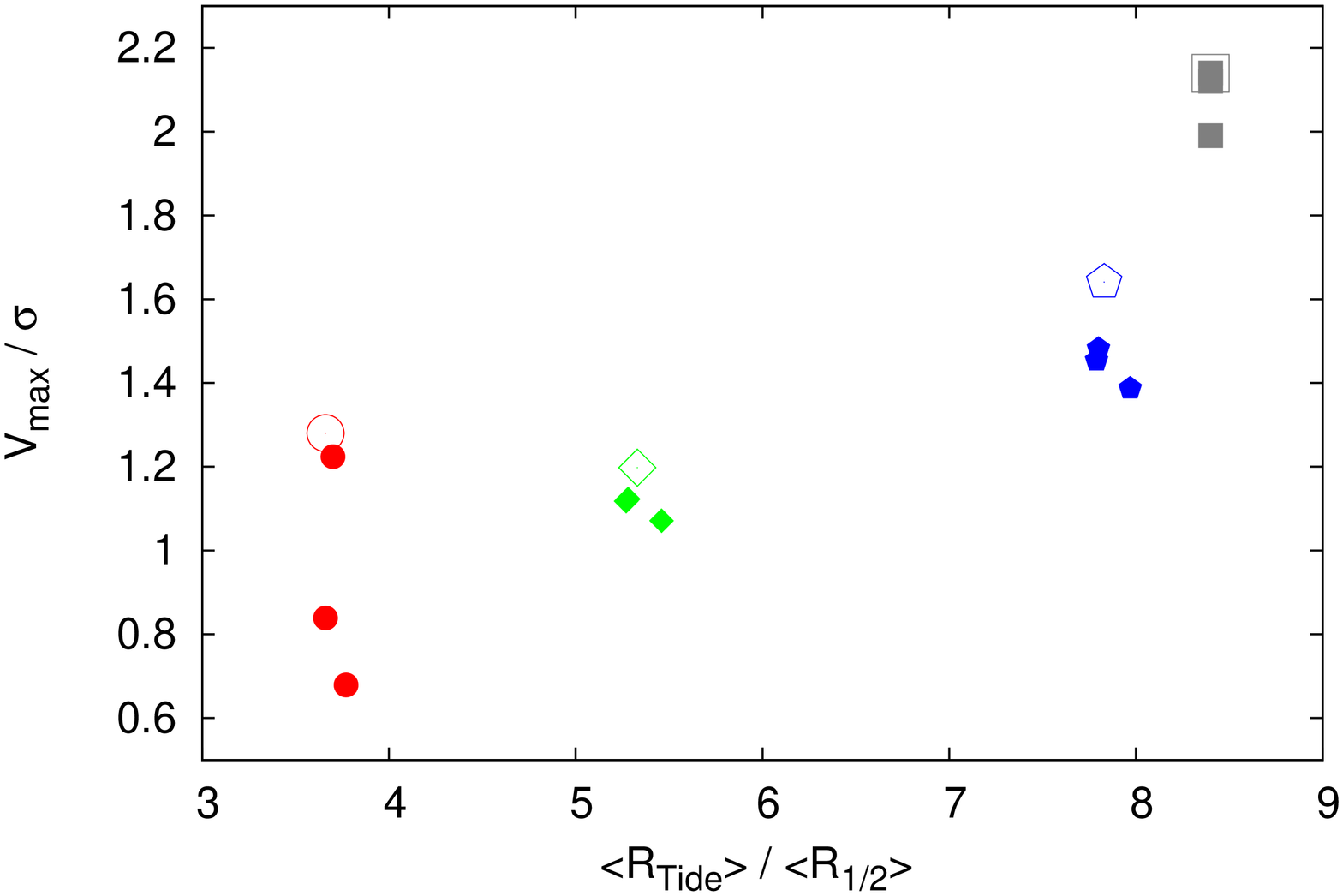}
   \caption{\emph{Top:} Axis ratio of the stellar component of the galaxy remnants versus the averaged tidal radius in units of the averaged half-mass radius.  
               \emph{Bottom:} Ratio of the rotational peak velocity and the mean velocity dispersion of the galaxy remnants versus the averaged tidal radius in units of the averaged half-mass radius of the stellar component. Both panels only contain simulations without any inclination between orbital plane and galactic disk. Open symbols represent simulations without encounters with other cluster members.} 
   \label{Resultieren}
   \end{figure}

The top panel of Fig.~\ref{Resultieren} shows the thickness of the galaxy remnant against the ratio of the average tidal radius to the average half-mass radius. The thickness of the galaxies is measured as the ratio of  the square root of the quadratic mean height of stars and the square root of the quadratic mean radius of stars ($ c/a = \sqrt{ \left\langle z^2 \right\rangle} / \sqrt{ \left\langle r^2 \right\rangle } $). When averaging over the runs of each setup, we find that the stronger the tidal force, the stronger the average thickening of the galaxies. At a fixed ratio of averaged tidal radius to averaged half-mass radius, the thickening was increased by the individual encounters with cluster members. Within a given setup, the stronger the stellar mass loss induced by such an encounter, the stronger the induced thickening. But also without additional mass loss, encounters were able to disturb the disk and induce a small thickening of the galaxy. 
The bottom panel of Fig.~\ref{Resultieren} shows the ratio of maximal rotation velocity and the mean velocity dispersion of the stellar component. The galaxy model started with a rotation-dominated stellar component, but the rotational support decreased with time by mass stripping and disturbances. Even if  only small or no additional stellar mass loss happened -- as in the case of the simulations with an apocentre of 1.5 Mpc -- the ratio of peak velocity and mean dispersion could be decreased by perturbations by $\sim 10\%$ or more. In the two simulations where the stellar mass loss was above $30\%$,  the ratio of the peak velocity and the mean velocity dispersion drops below one and the systems become dominated by pressure.

\subsection{Effect of initial structural parameters}

Changing the initial extent and the mass of a galaxy model is another way to change the ratio of the averaged tidal radius and the averaged half-mass radius of the stellar component while keeping the orbital parameters fixed.  For a better comparison we plot the results of the simulations with the small model, together with those of section 3.1, where we used the large model in figures \ref{MstarOrb}, \ref{M_Star-R_tide_onlyOrbite2}, and \ref{Resultieren}. It illustrates how important the initial structural parameters of a galaxy model are. They are as relevant as the orbital parameters. 

In the simulation without encounters, the small model lost no stellar mass, experienced no thickening, its ratio of peak velocity and mean dispersion decreased only by $\sim 6\%$ and only $\sim 6\%$ of its dark matter mass got lost. The small model also had a smaller cross section for encounters with other cluster members, which made it hard to be influenced by such encounters. Only one encounter in one of the runs was strong enough to disturb the galaxy, and it took only $2\%$ of the stellar mass away. However, this single encounter increased the thickness of the disk by a factor of two and decreased the ratio of maximal rotation velocity and mean dispersion by an additional $8\%$. These results fit well with the finding of sections 3.1 and 3.2 that the extent of a galaxy has to be close to its tidal radius and the galaxy has to be in a region with high galaxy density to get tidally transformed efficiently.

\subsection{Effect of  inclination}

   \begin{figure}
   \centering
   \includegraphics[angle=-90. ,width=0.5\textwidth]{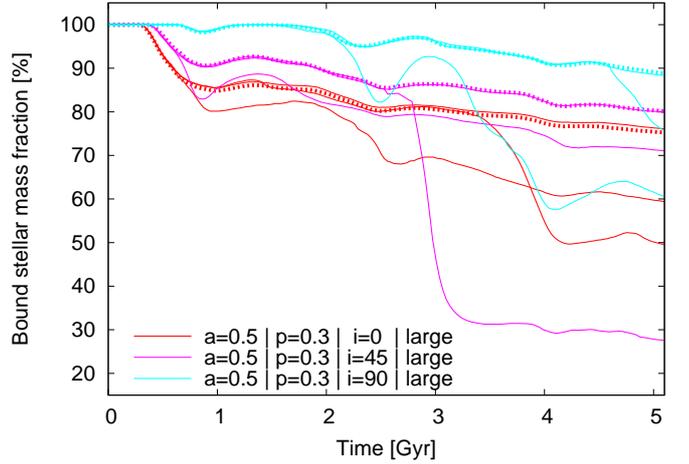}
   \caption{Bound stellar mass fraction in the simulations with different orientations of the galactic disk. The dotted lines represent simulations without encounters with other cluster members.}
   \label{MstarInc}
   \end{figure}

   \begin{figure}
   \includegraphics[angle=-90. ,width=0.5\textwidth]{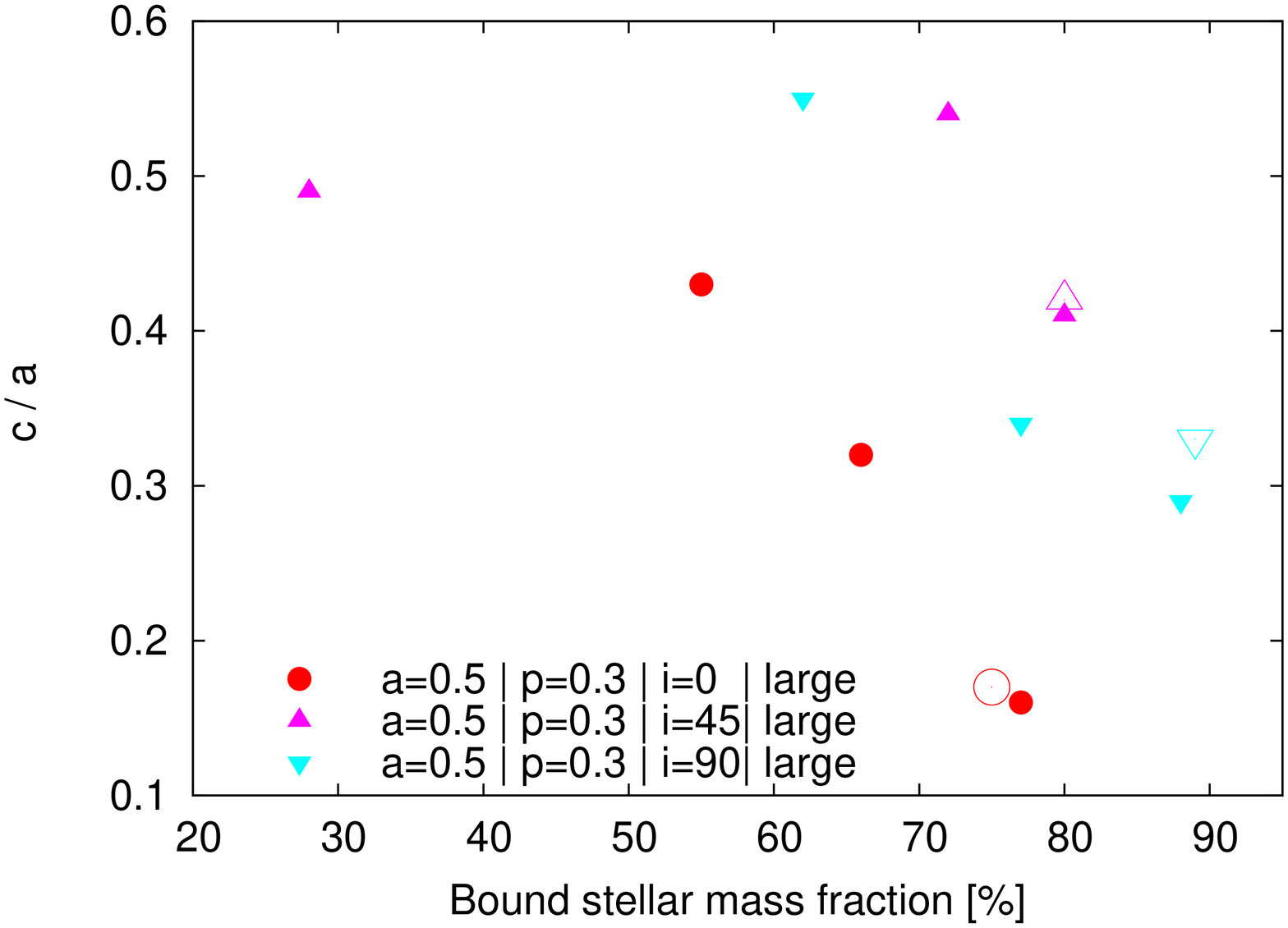}  
   \includegraphics[angle=-90. ,width=0.5\textwidth]{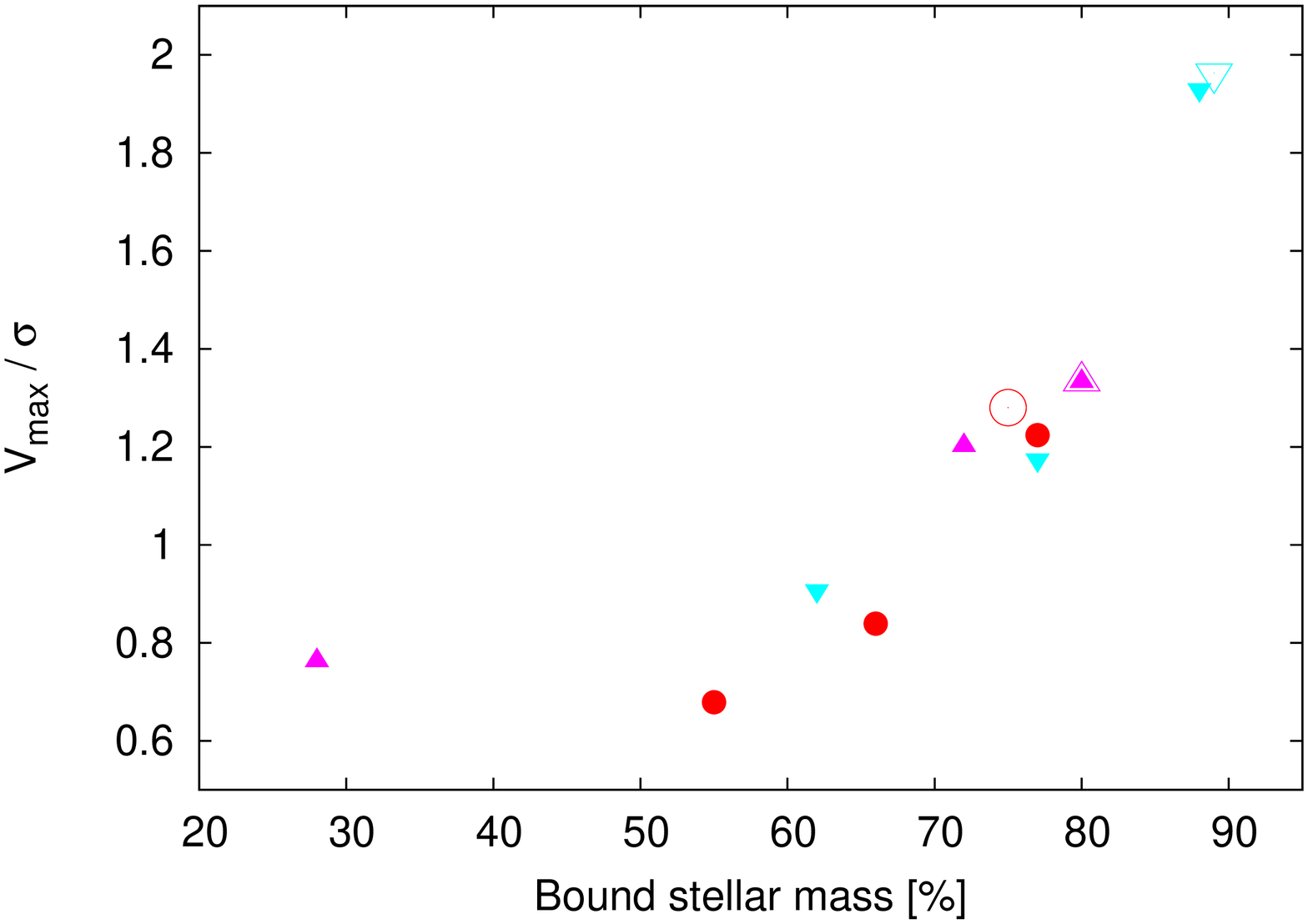}
   \caption{As in Fig.~\ref{Resultieren}, but for simulations with an inclination between orbital plane and galactic disk.} 
   \label{Resultieren-Inc}
   \end{figure}

The inclination of the galactic disk relative to the orbital plane is an additional parameter that influences the efficiency of the tidal transformation of galaxies. We ran simulations for galaxies with different inclinations ($0^{\circ}$, $45^{\circ}$, and $90^{\circ}$) on those of our orbits that are farthest inside. An inclination is expected to decrease the efficiency of mass stripping by the global tidal field \citep{2012MNRAS.424.2401V}, but in general it is not expected to change the effect of close encounters. Figure~\ref{MstarInc} shows the evolution of the bound stellar mass fraction in the simulations with different inclinations and the closest orbit. As expected the mass loss induced by the global tidal field alone (dotted lines in Fig. ~\ref{MstarInc}) decreases with increasing inclination.  In the case of a perpendicular orientation of the disk ($90^{\circ}$), the stellar mass loss induced by the global tidal field  was reduced by half compared to the case where the disk is parallel to the orbital plane ($0^{\circ}$).  
The additional mass loss caused by encounters with other cluster members still shows a large scatter. 

Figure~\ref{Resultieren-Inc} shows the thickness and the ratio of peak velocity to the mean velocity dispersion against the bound stellar mass fraction. Galaxies with the same stellar mass loss show stronger thickening if they do not move parallel to their disk. Under the influence of the global tidal field and without encounters, the galaxies with an inclination thicken more. Especially with an inclination of $45^{\circ}$, the galaxy thickens by a factor of five, whereas in the case of $0^{\circ}$, the galaxy thickens only by a factor of two. The simulations without encounters show that the galaxy with a perpendicular orientation of the disk to the orbital plane could preserve its rotational support much better than the galaxies with other inclinations. Nevertheless, under the influence of encounters, a galaxy with perpendicular orientation of the disk also became pressure-dominated. This highlights the stochastical nature of the harassment scenario again. 

One has to stress that the effects an inclination has on the interaction with the global tidal field of a cluster are diverse:  it could reduce the mass loss, amplify the thickening, and reduce the kinematical transformation. However, it seems that the influence on each of these quantities differs between the different orientations. Therefore, to describe the evolution of galaxies in a cluster environment, one has to pay attention to the different possible orientations of the disk.

\section{Comparison with observations}

   \begin{figure*}
   \centering
   \includegraphics[angle=-90. ,width=0.9\textwidth]{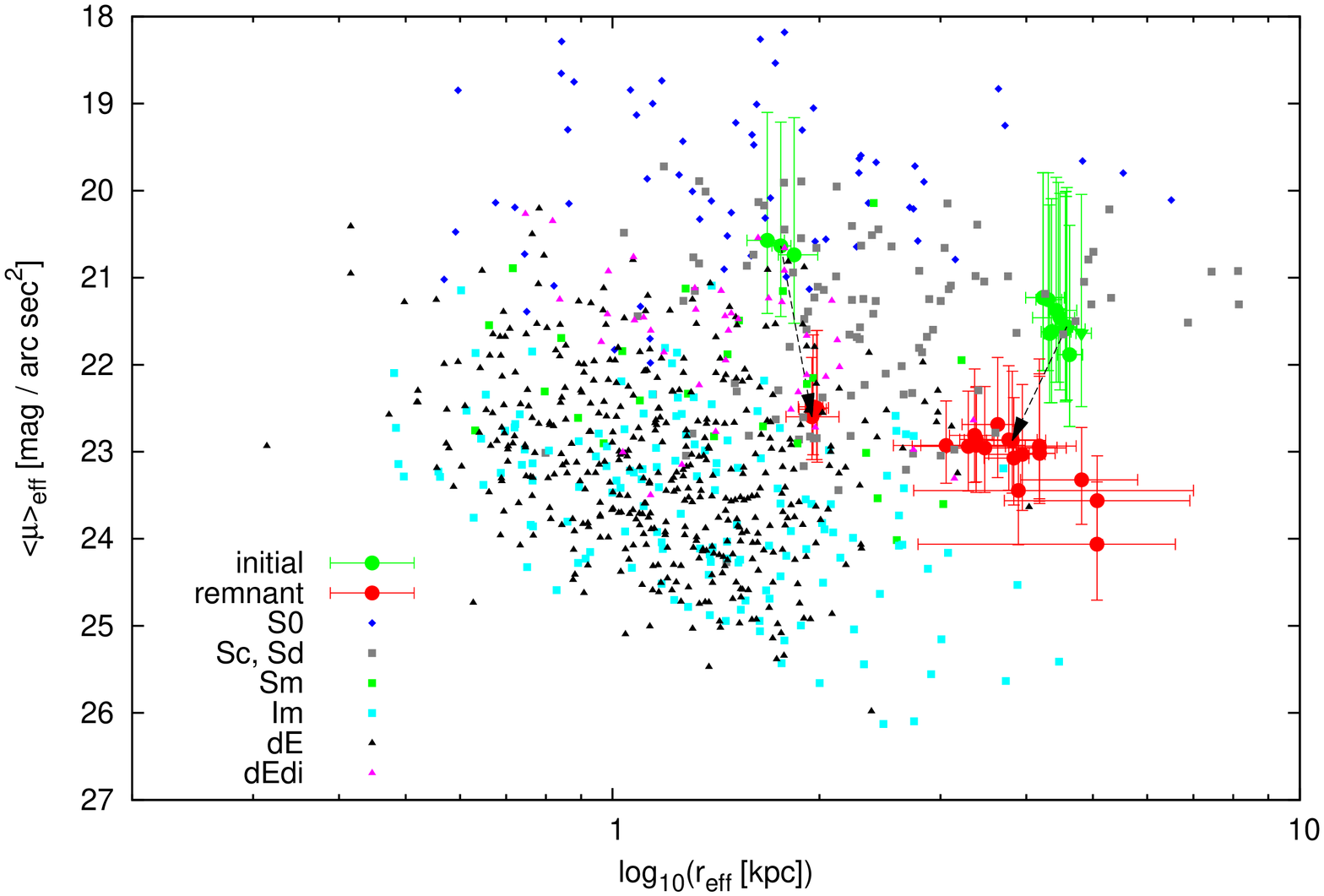}

   \includegraphics[angle=-90. ,width=0.9\textwidth]{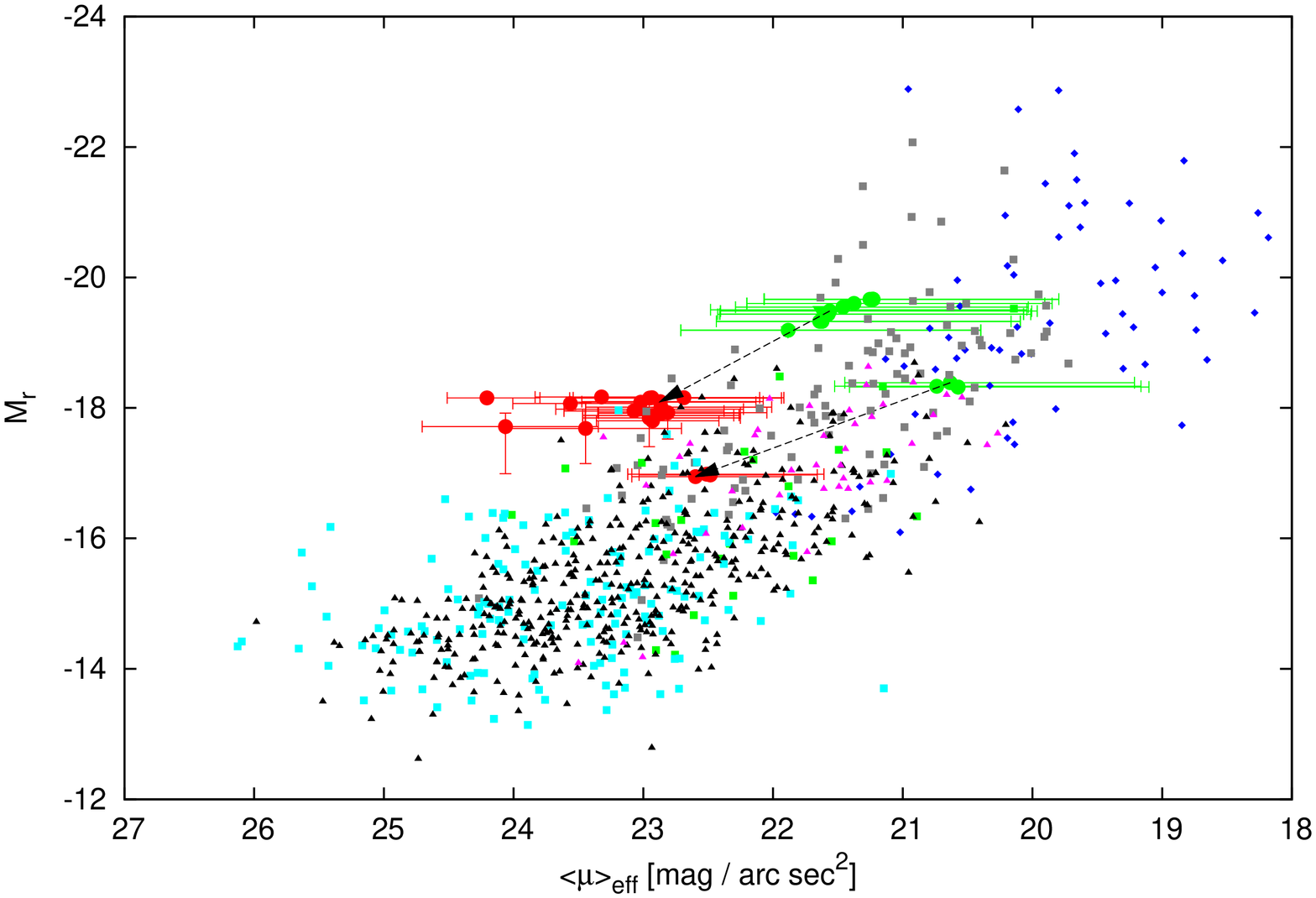}
   
   \caption{\emph{Top:} Distribution of Virgo galaxies, as well as the initial galaxy models and remnants in the $\rm \left\langle \mu \right\rangle_{\rm eff}  - r_{\rm eff}$ plane.
           \emph{Bottom:} Distribution of Virgo galaxies, as well as the initial galaxy models and remnants in the $\rm M_r - \left\langle \mu \right\rangle_{\rm eff}$ plane.  
                      The points for the initial and remnant galaxies are the mean values averaged over different projection angles (edge on = $90^{\circ}$,$67.5^{\circ}$,$45^{\circ}$,$22.5^{\circ}$, face on = $0^{\circ}$) along the x-axis and along the y-axis. The error bars span over the range from the minimum to the maximum value of the projections.}
   \label{ReffSurface}
   \end{figure*}
%

Figure~\ref{ReffSurface} shows the distribution of different galaxy types of the Virgo cluster \citep{2007ApJ...660.1186L,2008ApJ...689L..25J,2009AN....330..948J, 2014A&A...562A..49M} and the distribution of our initial galaxy models (green), as well as their remnants (red) in the $ \left\langle \mu \right\rangle_{\rm eff}  - r_{\rm eff}$ (top panel) and in the $ M_r - \left\langle \mu \right\rangle_{\rm eff}$ (bottom panel) plane. The points for the initial and remnant galaxies are the mean values averaged over different projection angles (edge-on = $90^{\circ}$, $67.5^{\circ}$, $45^{\circ}$,$22.5^{\circ}$, $0^{\circ}$ = face-on) along the x-axis and along the y-axis; the error bars span the range from the minimum to the maximum value of these projections of the galaxies. 
The  small scatter of the initial models originates in the different random-number seeds that are used to assign masses and ages to the stellar particles. The large error bars in the surface brightness of the initial models stems from the fact that the surface brightness of the edge-on projection of the model is nearly 2 $\rm mag/arc sec^2$ brighter than the face-on projection.
The point clouds of initial and remnant galaxies that belong to simulations with the large model can be identified by their clear separation in effective radius and magnitudes in comparison to the point clouds of the smaller model. 

Compared to the distribution of the Sc/Sd galaxies in the Virgo cluster as given by \cite{2007ApJ...660.1186L}, \cite{2008ApJ...689L..25J, 2009AN....330..948J}, and \cite{2014A&A...562A..49M}, the initial large models that are based on the Sc/Sd parameters of \cite{2008MNRAS.388.1708G} tend to have larger radii. The initial small models were  smaller than the typical extent of the Virgo Sc/Sd galaxies, which have a mean effective radius of $\rm \overline{r_{eff}}$ = 2.63 kpc.
The point cloud of remnants of the large model overlaps partially with the more extended dEs and dEdis. But on average the remnants of the large model tend to have radii that are too large to fit with the distribution of dEs  well. Their magnitudes are close to -18 mag in $r$, which is often used as the threshold for classifying bright dwarf galaxies, but in general they are too bright for ordinary dEs or dEdis. This trend indicates that the stripping of stellar material, hence the truncation of the disk and the dimming of light in simulations with the large model, was not efficient enough to produce typical dE or dEdi galaxies. However, the direction of evolution of the large model implies that harassment of galaxies could potentially produce dE and dEdis if the effects were amplified in some way. 

The remnants of the small model are all placed inside the distribution of bright dE and dEdi galaxies. But the dimming of light of these remnants was only caused by the truncation of the star formation at the beginning of the simulation, not by stripping of stellar matter. We have to stress that it has been shown before that these galaxies were nearly unaffected by the environment, so these remnants are still disk galaxies with a small thickness (c/a = 0.09 - 0.19) and are strongly dominated by rotation ($v/\sigma$ = 1.99 - 2.14). They could not be used to explain the distribution of dE galaxies because they did not undergo a morphology transformation. Nonetheless, we can imagine that if the effect of the tidal interaction was amplified, then the direction of evolution of these galaxies would be similar to that of the large model, and they would undergo a morphological transformation and end up as dE galaxies. 

This comparison shows that there is already a partial overlap of harassed galaxies  with dE/dEdi galaxies after an evolution of 5 Gyr in a cluster environment. From these results one can expect that if the environmental effects were amplified, the harassment scenario could be one production channel of dE and dEdi galaxies.

\section{Discussion}

\subsection{Orbital parameters}

By varying the apocentre of the orbit of an infalling galaxy between 0.5 Mpc and 1.5 Mpc at a fixed pericentre of 0.3 Mpc, it turned out that the tidally induced morphological transformation of low-mass late-type galaxies to bright dE galaxies became more efficient, the closer the orbit of the galaxies was. With decreasing apocentre, the mass loss, the thickness, and the pressure support increased \citep[also see][]{2005MNRAS.364..607M,2010MNRAS.405.1723S}. However, the comparison of the galaxy remnants with observations of dE galaxies shows that even on the innermost orbit, the environmental effects were still not efficient enough to transform the infalling galaxies into typical dE galaxies within an evolution time of 5~Gyr. In general, the remnants are still too extended and too bright, which indicates that a stronger environment would be necessary to strip more material and complete the transformation, or the progenitor galaxies would have to be late-type disks of lower luminosity and size than our model \citep[see][]{GrahamJerjenGuzman2003}.

If one takes dynamical friction into account, galaxies that enter a cluster will sink deeper into the cluster core. We neglected this effect in our simulations and probed what happened to galaxies that did not end up near the cluster centre. This is a relevant question because one observes dE galaxies not only in cluster cores but also in the outskirts of clusters, so a main production process of dE galaxies would also have to work in these regions and not just in the very cluster centre.  

On our more extended orbit (apocentre 1.5 Mpc) in the outskirts of the cluster, the tidal forces of the cluster were not strong enough and the galaxy encounters not frequent enough to harass the galaxies. On this orbit, the galaxies passed the pericentre only once in 5 Gyr, and the mass loss, as well as the thickening of the galaxies, was marginal. These results agree with the study of \cite{2010MNRAS.405.1723S}, who found in their simulations that low-mass dwarf irregular galaxies that enter a galaxy cluster from its present-day virial radius were mostly unaffected by the tidal forces of the cluster or by interactions with outer cluster members. Only on those orbits that are very close to the cluster centre did \citeauthor{2010MNRAS.405.1723S} find the tidal interactions to be efficient to transform dwarf irregular galaxies. They conclude that galaxies that are transformed in a harassment scenario have to have entered a galaxy cluster a long time ago to reach the innermost regions of a galaxy cluster and get transformed. 

\cite{2012MNRAS.423.1277D} have shown by using the data of the Millennium simulation \citep{2005Natur.435..629S} that galaxies with halo masses below $\rm 10^{11} M_{\odot}$ that are in a region closer than $0.5 R_{\rm vir}$ today became  satellites of bigger systems 9~Gyr ago (median value). Nine billion years are nearly double the time that we simulated. However, we could expect from our simulation results that a galaxy that spends such a long time in an environment like a cluster core could complete the transformation to a dE galaxy. 

One point that needs to be taken into account when simulating such a long time period is that the cluster itself will evolve on this time scale. A galaxy that enters a cluster nine billion years ago enters a protocluster environment with a mass that is more like present-day groups.  \cite{2012MNRAS.424.2401V} have done simulations of the evolution of galaxies in a group-like tidal field to study the relevance of preprocessing of galaxies in groups. They find that the group tidal field can induce a morphological transformation \citep[also see][]{Mayer2001}. Taking the preprocessing in galaxy groups into account could thus also be an answer for how harassed galaxies could populate the outskirts of clusters, because not all protocluster members will end up in the cluster core. 

\cite{2007ApJ...660.1186L}  mention that the different infall times and the associated cluster centric distances could be an explanation of the origin of different dE subpopulations like dEdis and could give a natural answer to the question of why the different subpopulations of dEs follow a morphology-density relation. In this sense one could  interpret dEdis not only as galaxies in an uncompleted transformation phase \citep{2005MNRAS.364..607M} but also as a separate subpopulation that evolved largely in a different environment than ordinary dEs. In the same sense, all the different subpopulations of dEs may have evolved in parallel in different environments and would then be the present-day product of this parallel evolution \citep{2013MNRAS.432.1162L}.   

The next step in clarifying the relevance of the harassment scenario as a production channel of dE galaxies would have to be studies that simulate a longer period in time and take the evolution of the environment itself into account.

\subsection{Structural parameters}

We have probed the influence of structural parameters of the progenitor galaxy by using two different models: one with a small and one with a large disk component, which differ by a factor 2.5 in effective radius. While the size of the large model becomes smaller by $26 \%$ on average, the size of the small model remains unchanged (see Table~\ref{table:3}). The small model was sitting so deep inside its tidal radius that it remained nearly untouched by the global tidal field.
The slight increase in the half-light radius (Fig.~\ref{ReffSurface}) is due to the older stellar population of the bulge component.

Small galaxies are much harder to affect by the tidal force of the cluster or by encounters with other cluster members, in agreement with the findings of \cite{1999MNRAS.304..465M}. They studied the efficiency of the harassment scenario for low and high surface brightness galaxies and found that high surface brightness galaxies are very resistant to environmental effects. 

A galaxy like the small model has to be on an orbit closer to the cluster centre to undergo a similar evolution as the large model. To reach the same ratio of mean tidal radius and half-mass radius,  the small model would have to be about two times closer to the cluster centre than the closest orbit we  used for the large model if we assume that the enclosed cluster mass stays the same. If one takes the radial mass profile of the cluster into account, one can in fact estimate that the small model would have to be about seven times closer to the cluster centre. This means it would have to be on an orbit with an apocentre closer than 100 kpc to the cluster centre. This emphasizes again that the initial structural parameters of the progenitor galaxy are as relevant as the orbit of the galaxy.

\subsection{Inclination}

The inclination of the disk of an infalling galaxy has been found to be an important parameter for the influence of the cluster's tidal field.
The induced stellar mass loss was reduced by half when the disk was perpendicular to the orbital plane as compared to a parallel orientation on the same orbit. 
On the other hand, the thickening of galaxies with an inclination was stronger than for parallel oriented galaxies. Especially with an inclination of $45^{\circ}$, the thickening induced by tidal forces was more than twice as large as the parallel oriented galaxy. Additionally, it turned out that galaxies with a perpendicular orientation could shield themselves better against kinematical transformations by the tidal field. 

\cite{2012MNRAS.424.2401V} have found in their simulations of the evolution of galaxies in a group-like tidal field that galaxies that are initially face-on or retrograde-oriented lost less mass and preserved their disk structure and kinematic for a longer time than prograde galaxies. \footnote{Our simulations only included prograde rotation.} They conclude that this could be an explanation for disk galaxies in the inner regions of groups and clusters. We agree with \cite{2012MNRAS.424.2401V} that galaxies with an inclination experienced less mass loss and kinematical transformations. 

We speculate that the effect of inclination could be an origin for dEdi galaxies. dEdis appear to be less round than ordinary dEs \citep{2006AJ....132..497L}, but with a thickness of 0.35 - 0.4, they are somewhat thicker than typical late-type disk galaxies at similar stellar mass \citep[cf.][]{SanchezJanssen2010}. The rotation parameter $v_{max}/\sigma$ of dEdi galaxies is smaller than for disk galaxies but higher than for ordinary dE galaxies \citep{2011A&A...526A.114T,Toloba2014p3}. We found that our simulated galaxies with an inclination of $45^{\circ}$ or $90^{\circ}$ thicken more and reach the regime of dEdi galaxies. Galaxies with a disk perpendicular to their orbital plane ($90^{\circ}$) could shield their kinematics much better against the influence of the tidal field. Both observations together lead to the speculation that disk galaxies that enter a cluster at a high inclination $45^{\circ}$ - $90^{\circ}$ have a good chance of ending up with a thick disk with high $v_{max}/\sigma$ values. In that case, harassed high-inclination disk galaxies may be an origin for dEdi galaxies.

\section{Conclusions}

Harassment of galaxies is a complex process. Its efficiency depends on a combination of many parameters, such as the orbit, the inclination, and the structural parameters of the progenitor galaxy, as well as on the properties of the galaxy cluster itself. It is likely that harassment is one of the most important engines of galaxy evolution inside galaxy clusters, and is often referred to in past studies. However, this process would have had to start at a very early epoch in protocluster environments to complete the transformation by today. To fully understand the overall importance of harassment, more simulations that reproduce the complicated interaction between galaxies and their environment are needed.

\begin{acknowledgements}

We thank Isabel Franco for her efforts with a previous simulation study that eventually led to the project presented here.

D.B.\ acknowledges funding by the IMPRS for Astronomy \& Cosmic Physics at the University of Heidelberg.
D.B.\ and R.S.\ acknowledge support by Chinese Academy of Sciences through the Silk Road Project at NAOC, through the Chinese Academy of Sciences Visiting Professorship for Senior International Scientists, grant number $2009S1-5$ (R.S.), and through the ``Qianren''special foreign experts program of China.
T.L.\ was supported within the framework of the Excellence Initiative by the German Research Foundation (DFG) through the Heidelberg Graduate School of Fundamental Physics (grant number GSC 129/1).
C.O.\  appreciates funding by the German Research Foundation (DFG) grant
OL 350/1-1 and SFB~881.
The work of C.O.\ and R.S.\ was supported in part by the National Science Foundation under grant PHYS-1066293 and by the hospitality of the Aspen Center for Physics.
R.K. \ gratefully acknowledges financial support from STScI theory grant HST-AR-12840.01-A. Support for Program number HST-AR-12840.01-A was provided by NASA through a grant from the Space Telescope Science Institute, which is operated by the Association of Universities for Research in Astronomy, Incorporated, under NASA contract NAS5-26555.

We used computing resources at the Center of Information and Computing at National Astronomical Observatories, Chinese Academy of Sciences, funded by the Ministry of Finance of the People's Republic of China under grant $ZDYZ2008-2$, and at the ZAH, Heidelberg University, GPU clusters {\tt hydra} and {\tt kepler}, funded under grants I/80041-043 and I/84678/84680 of the Volkswagen Foundation.

\end{acknowledgements}

\appendix
\section{Equations of motions}

The reference frame that was used to describe the equations of motion of a particle in our simulation was centred on the density peak of the bulge. The galaxies themselves moved on an eccentric orbit around the cluster centre. Additionally, we rotated our coordinate system in such a way that the x-axis of the reference frame always pointed away from the centre of the cluster. This means that our coordinate system is not an inertial system, but an accelerated and rotated one. 

The equation of motion of a particle is 
\begin{eqnarray}
\vec{\ddot{r}} = \vec{f_{\rm gal}} + \vec{f_{\rm cl}} + \vec{a_0} + \vec{f_{\rm ce}} + \vec{f_{\rm co}} + \vec{f_{\rm eu}}   
.\end{eqnarray}
The first two terms are the gravitational forces of the galaxy $\vec{f_{\rm gal}}$ and the cluster $\vec{f_{\rm cl}}$, which acts on a particle. The gravitational force of the galaxy was calculated by direct summation of the two-body interaction of the particles, and the gravitational force of the cluster was calculated by an analytical force field as described in section 2.1:
\begin{eqnarray}
\vec{f_{\rm gal}} = - \vec{\nabla} \Phi_{\rm gal} \\
\vec{f_{\rm cl}} \ = - \vec{\nabla} \Phi_{\rm cl} \ \
.\end{eqnarray}
The third term of equation A.1 is the force of inertia $\vec{a_0}$. It has the same strength but inverse direction as the force that accelerates the galaxy on its orbit around the cluster centre. The distance of the cluster centre and the origin of ordinates is $r_0$. The direction of the force of inertia points away from the cluster centre, which means that in our rotating reference frame along the direction of the x-axis, that 
\begin{eqnarray}
\vec{a_0} =  \left[ \vec{e_x} \frac{d \Phi_{\rm cl}}{d r_{\rm cl}} \right]_{r_0} 
.\end{eqnarray}
Since it is a rotating reference frame, we have to take the centrifugal $\vec{f_{\rm ce}}$, Coriolis $\vec{f_{\rm co}}$ and Euler force $\vec{f_{\rm eu}}$ into account. The orbital plane of the galaxy within the cluster is chosen so that the angular velocity of the motion of the galaxy around the cluster centre is pointing in the z direction. This angular velocity is the same as the angular velocity that is used to rotate the reference frame:
\begin{eqnarray}
\vec{f_{\rm ce}} = -\vec{\omega} \times \left(  \vec{\omega} \times \vec{r} \right) = \omega^2 \left( x \vec{e_x} + y \vec{e_y} \right) \ \ \\
\vec{f_{\rm co}} = -2 \vec{\omega} \times \dot{\vec{r}} = 2\omega  \left(\dot{y} \vec{e_x} - \dot{x} \vec{e_y} \right)  \ \ \ \ \ \ \ \ \  \\
\vec{f_{\rm eu}} = - \dot{\vec{\omega}} \times \vec{r} = \dot{\omega} \left( y \vec{e_x} - x \vec{e_y} \right) \ \ \ \  \ \ \ \ \ \ \ \ \ \
.\end{eqnarray}
After an expansion of the cluster potential in a Taylor series to the first order around the origin of ordinates, the equations of motions are 
\begin{eqnarray}
\ddot{x} \approx - \frac{\partial}{\partial x} \Phi_{gal} + \left( \omega^2 -\left[ \frac{\partial^2 \Phi_{cl} }{\partial r_cl^2} \right]_{r_0} \right) x + 2 \omega \dot{y} +\dot{\omega} y \ \ \ \\
\ddot{y} \approx - \frac{\partial}{\partial y} \Phi_{gal} + \left( \omega^2 -\left[ \frac{\partial \Phi_{cl} }{\partial r_cl} \right]_{r_0} \frac{1}{r_0} \right) y - 2 \omega \dot{x} - \dot{\omega} x \\
\ddot{z} \approx - \frac{\partial}{\partial y} \Phi_{cl} - \left[ \frac{\partial \Phi_{cl}}{\partial r_{cl}} \right]_{r_0} \frac{z}{r_0} \ \ \ \  \ \ \ \ \ \ \ \ \ \ \ \  \ \ \ \ \ \ \ \ \ \ \ \ \ \ \ \ \ \ \ \  
.\end{eqnarray}
If one assumes a circular orbit for the galaxy $ \dot{\omega} = 0, $ these equations of motion simplify to the Hill's approximation.

\section{Properties of encounter galaxies}

The typical time between two encounters depends on the local galaxy density n(r) and the local velocity dispersion $\sigma(r)$ of the cluster, which are functions of the cluster centric distance r. We take all those encounters into account that pass the infalling galaxy within a distance of $p_{max} = 60 kpc$:

\begin{eqnarray}
\tau = \left( \Sigma  \ \  n(r)  \ \  \sigma(r)  \right)^{-1}\quad,
\end{eqnarray}
where $\Sigma$ is the geometrical cross section of such a galaxy-galaxy encounter $\Sigma = \pi p_{max}^{2} $ . 
The local galaxy density was derived by assuming a galaxy distribution that follows a power law:
\begin{eqnarray}
\frac{dN}{dr} = N_0 \ \ \ r^{\beta}
.\end{eqnarray}
The parameters $N_0 = 154$ and $\beta = 0.6$ were determined by a fit to the radial galaxy distribution of those galaxy clusters of the Millennium II simulation \citep{2009MNRAS.398.1150B} that have a mass between $2.4 - 4 \cdot 10^{14} M_{\odot}$. 

Cluster members with total masses down to $0.1 M_{gal}$ were taken into account, where $M_{gal}$ was the mass of the infalling galaxy. The total masses of the galaxies were taken from the corresponding semi-analytic model of \citep{2011MNRAS.413..101G}.
The local velocity dispersion was calculated numerically as a solution of the spherical Jeans-equation for the cluster potential. 
The occurrence of an encounter was simulated by randomly placing the particle that represents the perturber galaxy on a sphere with a radius of 200kpc around the infalling galaxy. The velocities of the flying-by galaxies followed a Maxwell-Boltzmann distribution. The velocity vector of the flying-by galaxy pointed to the centre of the sphere, but offset by the impact parameter of the encounter.
The impact parameter was chosen randomly based on the geometrical cross section, so that the probability of an impact parameter $p \in [p_{min},p_{max}]$ is given by

\begin{eqnarray}
f(p) =  2 \frac{p dp}{\left( p_{max}^2 - p_{min}^2 \right) }
.\end{eqnarray}

One also has to consider that the flying-by galaxy will by focused on the infalling galaxy by gravitation. We corrected the offset to take this gravitational focusing into account by following \cite{2009ApJ...697..458S}.

The masses of the perturber galaxies were chosen randomly by following  the mass distribution of the aforementioned galaxy clusters of the Millennium II simulation \citep{2009MNRAS.398.1150B}. To determine the mass distribution, a power-law ansatz was used:

\begin{eqnarray}
 \frac{dN}{dm} = N_0 \ \ \ m^{-\alpha} 
.\end{eqnarray}

The fit yielded $\alpha = 2.0$

\bibliographystyle{aa} 
\bibliography{25235}

\begin{thebibliography}{73}
\expandafter\ifx\csname natexlab\endcsname\relax\def\natexlab#1{#1}\fi

\bibitem[{{Aarseth}(1999)}]{1999PASP..111.1333A}
{Aarseth}, S.~J. 1999, PASP, 111, 1333

\bibitem[{{Aarseth}(2003)}]{Aarseth2003}
{Aarseth}, S.~J. 2003, {Gravitational N-Body Simulations}

\bibitem[{{Adelman-McCarthy} {et~al.}(2006){Adelman-McCarthy}, {Ag{\"u}eros},
  {Allam}, {Anderson}, {Anderson}, {Annis}, {Bahcall}, {Baldry}, {Barentine},
  {Berlind}, {Bernardi}, {Blanton}, {Boroski}, {Brewington}, {Brinchmann},
  {Brinkmann}, {Brunner}, {Budav{\'a}ri}, {Carey}, {Carr}, {Castander},
  {Connolly}, {Csabai}, {Czarapata}, {Dalcanton}, {Doi}, {Dong}, {Eisenstein},
  {Evans}, {Fan}, {Finkbeiner}, {Friedman}, {Frieman}, {Fukugita}, {Gillespie},
  {Glazebrook}, {Gray}, {Grebel}, {Gunn}, {Gurbani}, {de Haas}, {Hall},
  {Harris}, {Harvanek}, {Hawley}, {Hayes}, {Hendry}, {Hennessy}, {Hindsley},
  {Hirata}, {Hogan}, {Hogg}, {Holmgren}, {Holtzman}, {Ichikawa}, {Ivezi{\'c}},
  {Jester}, {Johnston}, {Jorgensen}, {Juri{\'c}}, {Kent}, {Kleinman}, {Knapp},
  {Kniazev}, {Kron}, {Krzesinski}, {Kuropatkin}, {Lamb}, {Lampeitl}, {Lee},
  {Leger}, {Lin}, {Long}, {Loveday}, {Lupton}, {Margon},
  {Mart{\'{\i}}nez-Delgado}, {Mandelbaum}, {Matsubara}, {McGehee}, {McKay},
  {Meiksin}, {Munn}, {Nakajima}, {Nash}, {Neilsen}, {Newberg}, {Newman},
  {Nichol}, {Nicinski}, {Nieto-Santisteban}, {Nitta}, {O'Mullane}, {Okamura},
  {Owen}, {Padmanabhan}, {Pauls}, {Peoples}, {Pier}, {Pope}, {Pourbaix},
  {Quinn}, {Richards}, {Richmond}, {Rockosi}, {Schlegel}, {Schneider},
  {Schroeder}, {Scranton}, {Seljak}, {Sheldon}, {Shimasaku}, {Smith}, {Smol{\v
  c}i{\'c}}, {Snedden}, {Stoughton}, {Strauss}, {SubbaRao}, {Szalay},
  {Szapudi}, {Szkody}, {Tegmark}, {Thakar}, {Tucker}, {Uomoto}, {Vanden Berk},
  {Vandenberg}, {Vogeley}, {Voges}, {Vogt}, {Walkowicz}, {Weinberg}, {West},
  {White}, {Xu}, {Yanny}, {Yocum}, {York}, {Zehavi}, {Zibetti}, \&
  {Zucker}}]{2006ApJS..162...38A}
{Adelman-McCarthy}, J.~K., {Ag{\"u}eros}, M.~A., {Allam}, S.~S., {et~al.} 2006,
  \apjs, 162, 38

\bibitem[{{Aguerri} \&
  {Gonz{\'a}lez-Garc{\'{\i}}a}(2009)}]{2009A&A...494..891A}
{Aguerri}, J.~A.~L. \& {Gonz{\'a}lez-Garc{\'{\i}}a}, A.~C. 2009, \aap, 494, 891

\bibitem[{{Bah{\'e}} {et~al.}(2013){Bah{\'e}}, {McCarthy}, {Balogh}, \&
  {Font}}]{2013MNRAS.430.3017B}
{Bah{\'e}}, Y.~M., {McCarthy}, I.~G., {Balogh}, M.~L., \& {Font}, A.~S. 2013,
  \mnras, 430, 3017

\bibitem[{{Barazza} {et~al.}(2009){Barazza}, {Wolf}, {Gray}, {Jogee}, {Balogh},
  {McIntosh}, {Bacon}, {Barden}, {Bell}, {B{\"o}hm}, {Caldwell},
  {H{\"a}ussler}, {Heiderman}, {Heymans}, {Jahnke}, {van Kampen}, {Lane},
  {Marinova}, {Meisenheimer}, {Peng}, {Sanchez}, {Taylor}, {Wisotzki}, \&
  {Zheng}}]{2009A&A...508..665B}
{Barazza}, F.~D., {Wolf}, C., {Gray}, M.~E., {et~al.} 2009, \aap, 508, 665

\bibitem[{{Beasley} {et~al.}(2006){Beasley}, {Strader}, {Brodie}, {Cenarro}, \&
  {Geha}}]{Beasley2006}
{Beasley}, M.~A., {Strader}, J., {Brodie}, J.~P., {Cenarro}, A.~J., \& {Geha},
  M. 2006, \aj, 131, 814

\bibitem[{{Bertin} \& {Arnouts}(1996)}]{1996A&AS..117..393B}
{Bertin}, E. \& {Arnouts}, S. 1996, \aaps, 117, 393

\bibitem[{{Binggeli} {et~al.}(1985){Binggeli}, {Sandage}, \&
  {Tammann}}]{1985AJ.....90.1681B}
{Binggeli}, B., {Sandage}, A., \& {Tammann}, G.~A. 1985, \aj, 90, 1681

\bibitem[{{Binggeli} {et~al.}(1987){Binggeli}, {Tammann}, \&
  {Sandage}}]{1987AJ.....94..251B}
{Binggeli}, B., {Tammann}, G.~A., \& {Sandage}, A. 1987, \aj, 94, 251

\bibitem[{{Binggeli} {et~al.}(1990){Binggeli}, {Tarenghi}, \&
  {Sandage}}]{Binggeli1990}
{Binggeli}, B., {Tarenghi}, M., \& {Sandage}, A. 1990, \aap, 228, 42

\bibitem[{{Bizyaev} \& {Mitronova}(2009)}]{2009ApJ...702.1567B}
{Bizyaev}, D. \& {Mitronova}, S. 2009, \apj, 702, 1567

\bibitem[{{Boselli} {et~al.}(2008){Boselli}, {Boissier}, {Cortese}, \&
  {Gavazzi}}]{2008ApJ...674..742B}
{Boselli}, A., {Boissier}, S., {Cortese}, L., \& {Gavazzi}, G. 2008, \apj, 674,
  742

\bibitem[{{Boylan-Kolchin} {et~al.}(2009){Boylan-Kolchin}, {Springel}, {White},
  {Jenkins}, \& {Lemson}}]{2009MNRAS.398.1150B}
{Boylan-Kolchin}, M., {Springel}, V., {White}, S.~D.~M., {Jenkins}, A., \&
  {Lemson}, G. 2009, \mnras, 398, 1150

\bibitem[{{Chang} {et~al.}(2013){Chang}, {Macci{\`o}}, \&
  {Kang}}]{2013MNRAS.431.3533C}
{Chang}, J., {Macci{\`o}}, A.~V., \& {Kang}, X. 2013, \mnras, 431, 3533

\bibitem[{{Chen} {et~al.}(2010){Chen}, {C{\^o}t{\'e}}, {West}, {Peng}, \&
  {Ferrarese}}]{Chen2010}
{Chen}, C.-W., {C{\^o}t{\'e}}, P., {West}, A.~A., {Peng}, E.~W., \&
  {Ferrarese}, L. 2010, \apjs, 191, 1

\bibitem[{{C{\^o}t{\'e}} {et~al.}(2007){C{\^o}t{\'e}}, {Ferrarese},
  {Jord{\'a}n}, {Blakeslee}, {Chen}, {Infante}, {Merritt}, {Mei}, {Peng},
  {Tonry}, {West}, \& {West}}]{Cote2007}
{C{\^o}t{\'e}}, P., {Ferrarese}, L., {Jord{\'a}n}, A., {et~al.} 2007, \apj,
  671, 1456

\bibitem[{{De Lucia} {et~al.}(2012){De Lucia}, {Weinmann}, {Poggianti},
  {Arag{\'o}n-Salamanca}, \& {Zaritsky}}]{2012MNRAS.423.1277D}
{De Lucia}, G., {Weinmann}, S., {Poggianti}, B.~M., {Arag{\'o}n-Salamanca}, A.,
  \& {Zaritsky}, D. 2012, \mnras, 423, 1277

\bibitem[{{De Rijcke} {et~al.}(2003){De Rijcke}, {Dejonghe}, {Zeilinger}, \&
  {Hau}}]{DeRijcke2003}
{De Rijcke}, S., {Dejonghe}, H., {Zeilinger}, W.~W., \& {Hau}, G.~K.~T. 2003,
  \aap, 400, 119

\bibitem[{{De Rijcke} {et~al.}(2005){De Rijcke}, {Michielsen}, {Dejonghe},
  {Zeilinger}, \& {Hau}}]{DeRijcke2005}
{De Rijcke}, S., {Michielsen}, D., {Dejonghe}, H., {Zeilinger}, W.~W., \&
  {Hau}, G.~K.~T. 2005, \aap, 438, 491

\bibitem[{{Dressler}(1980)}]{1980ApJ...236..351D}
{Dressler}, A. 1980, \apj, 236, 351

\bibitem[{{Evans}(1993)}]{1993MNRAS.260..191E}
{Evans}, N.~W. 1993, \mnras, 260, 191

\bibitem[{{Gavazzi} {et~al.}(2002){Gavazzi}, {Bonfanti}, {Sanvito}, {Boselli},
  \& {Scodeggio}}]{2002ApJ...576..135G}
{Gavazzi}, G., {Bonfanti}, C., {Sanvito}, G., {Boselli}, A., \& {Scodeggio}, M.
  2002, \apj, 576, 135

\bibitem[{{Graham} {et~al.}(2003){Graham}, {Jerjen}, \&
  {Guzm{\'a}n}}]{GrahamJerjenGuzman2003}
{Graham}, A.~W., {Jerjen}, H., \& {Guzm{\'a}n}, R. 2003, \aj, 126, 1787

\bibitem[{{Graham} \& {Worley}(2008)}]{2008MNRAS.388.1708G}
{Graham}, A.~W. \& {Worley}, C.~C. 2008, \mnras, 388, 1708

\bibitem[{{Gunn} \& {Gott}(1972)}]{1972ApJ...176....1G}
{Gunn}, J.~E. \& {Gott}, III, J.~R. 1972, \apj, 176, 1

\bibitem[{{Guo} {et~al.}(2011){Guo}, {White}, {Boylan-Kolchin}, {De Lucia},
  {Kauffmann}, {Lemson}, {Li}, {Springel}, \& {Weinmann}}]{2011MNRAS.413..101G}
{Guo}, Q., {White}, S., {Boylan-Kolchin}, M., {et~al.} 2011, \mnras, 413, 101

\bibitem[{{Hurley} {et~al.}(2000){Hurley}, {Pols}, \&
  {Tout}}]{2000MNRAS.315..543H}
{Hurley}, J.~R., {Pols}, O.~R., \& {Tout}, C.~A. 2000, \mnras, 315, 543

\bibitem[{{Janz} \& {Lisker}(2008)}]{2008ApJ...689L..25J}
{Janz}, J. \& {Lisker}, T. 2008, \apjl, 689, L25

\bibitem[{{Janz} \& {Lisker}(2009)}]{2009AN....330..948J}
{Janz}, J. \& {Lisker}, T. 2009, Astronomische Nachrichten, 330, 948

\bibitem[{{Jerjen} {et~al.}(2000){Jerjen}, {Kalnajs}, \&
  {Binggeli}}]{2000A&A...358..845J}
{Jerjen}, H., {Kalnajs}, A., \& {Binggeli}, B. 2000, \aap, 358, 845

\bibitem[{{King}(1962)}]{1962AJ.....67..471K}
{King}, I. 1962, \aj, 67, 471

\bibitem[{{Kormendy}(1985)}]{Kormendy1985}
{Kormendy}, J. 1985, \apj, 295, 73

\bibitem[{{Kormendy} \& {Bender}(2012)}]{Kormendy2012}
{Kormendy}, J. \& {Bender}, R. 2012, \apjs, 198, 2

\bibitem[{{Kotulla} {et~al.}(2009){Kotulla}, {Fritze}, {Weilbacher}, \&
  {Anders}}]{2009MNRAS.396..462K}
{Kotulla}, R., {Fritze}, U., {Weilbacher}, P., \& {Anders}, P. 2009, \mnras,
  396, 462

\bibitem[{{Kroupa}(2001)}]{2001MNRAS.322..231K}
{Kroupa}, P. 2001, \mnras, 322, 231

\bibitem[{{Kuijken} \& {Dubinski}(1995)}]{1995MNRAS.277.1341K}
{Kuijken}, K. \& {Dubinski}, J. 1995, \mnras, 277, 1341

\bibitem[{{Lejeune} {et~al.}(1997){Lejeune}, {Cuisinier}, \&
  {Buser}}]{1997A&AS..125..229L}
{Lejeune}, T., {Cuisinier}, F., \& {Buser}, R. 1997, \aaps, 125, 229

\bibitem[{{Lejeune} {et~al.}(1998){Lejeune}, {Cuisinier}, \&
  {Buser}}]{1998A&AS..130...65L}
{Lejeune}, T., {Cuisinier}, F., \& {Buser}, R. 1998, \aaps, 130, 65

\bibitem[{{Lisker} {et~al.}(2006{\natexlab{a}}){Lisker}, {Glatt}, {Westera}, \&
  {Grebel}}]{2006AJ....132.2432L}
{Lisker}, T., {Glatt}, K., {Westera}, P., \& {Grebel}, E.~K.
  2006{\natexlab{a}}, \aj, 132, 2432

\bibitem[{{Lisker} {et~al.}(2006{\natexlab{b}}){Lisker}, {Grebel}, \&
  {Binggeli}}]{2006AJ....132..497L}
{Lisker}, T., {Grebel}, E.~K., \& {Binggeli}, B. 2006{\natexlab{b}}, \aj, 132,
  497

\bibitem[{{Lisker} {et~al.}(2007){Lisker}, {Grebel}, {Binggeli}, \&
  {Glatt}}]{2007ApJ...660.1186L}
{Lisker}, T., {Grebel}, E.~K., {Binggeli}, B., \& {Glatt}, K. 2007, \apj, 660,
  1186

\bibitem[{{Lisker} {et~al.}(2009){Lisker}, {Janz}, {Hensler}, {Kim}, {Rey},
  {Weinmann}, {Mastropietro}, {Hielscher}, {Paudel}, \&
  {Kotulla}}]{2009ApJ...706L.124L}
{Lisker}, T., {Janz}, J., {Hensler}, G., {et~al.} 2009, \apjl, 706, L124

\bibitem[{{Lisker} {et~al.}(2013){Lisker}, {Weinmann}, {Janz}, \&
  {Meyer}}]{2013MNRAS.432.1162L}
{Lisker}, T., {Weinmann}, S.~M., {Janz}, J., \& {Meyer}, H.~T. 2013, \mnras,
  432, 1162

\bibitem[{{Mastropietro} {et~al.}(2005){Mastropietro}, {Moore}, {Mayer},
  {Debattista}, {Piffaretti}, \& {Stadel}}]{2005MNRAS.364..607M}
{Mastropietro}, C., {Moore}, B., {Mayer}, L., {et~al.} 2005, \mnras, 364, 607

\bibitem[{{Mayer} {et~al.}(2001){Mayer}, {Governato}, {Colpi}, {Moore},
  {Quinn}, {Wadsley}, {Stadel}, \& {Lake}}]{Mayer2001}
{Mayer}, L., {Governato}, F., {Colpi}, M., {et~al.} 2001, \apj, 559, 754

\bibitem[{{McLaughlin}(1999)}]{1999ApJ...512L...9M}
{McLaughlin}, D.~E. 1999, \apjl, 512, L9

\bibitem[{{Meyer} {et~al.}(2014){Meyer}, {Lisker}, {Janz}, \&
  {Papaderos}}]{2014A&A...562A..49M}
{Meyer}, H.~T., {Lisker}, T., {Janz}, J., \& {Papaderos}, P. 2014, \aap, 562,
  A49

\bibitem[{{Moore} {et~al.}(1996){Moore}, {Katz}, {Lake}, {Dressler}, \&
  {Oemler}}]{1996Natur.379..613M}
{Moore}, B., {Katz}, N., {Lake}, G., {Dressler}, A., \& {Oemler}, A. 1996,
  \nat, 379, 613

\bibitem[{{Moore} {et~al.}(1998){Moore}, {Lake}, \&
  {Katz}}]{1998ApJ...495..139M}
{Moore}, B., {Lake}, G., \& {Katz}, N. 1998, \apj, 495, 139

\bibitem[{{Moore} {et~al.}(1999){Moore}, {Lake}, {Quinn}, \&
  {Stadel}}]{1999MNRAS.304..465M}
{Moore}, B., {Lake}, G., {Quinn}, T., \& {Stadel}, J. 1999, \mnras, 304, 465

\bibitem[{{Navarro} {et~al.}(1996){Navarro}, {Frenk}, \&
  {White}}]{1996ApJ...462..563N}
{Navarro}, J.~F., {Frenk}, C.~S., \& {White}, S.~D.~M. 1996, \apj, 462, 563

\bibitem[{{Nitadori} \& {Aarseth}(2012)}]{2012MNRAS.424..545N}
{Nitadori}, K. \& {Aarseth}, S.~J. 2012, \mnras, 424, 545

\bibitem[{{Pak} {et~al.}(2014){Pak}, {Rey}, {Lisker}, {Lee}, {Kim}, {Sung},
  {Jerjen}, \& {Chung}}]{Pak2014}
{Pak}, M., {Rey}, S.-C., {Lisker}, T., {et~al.} 2014, \mnras, in press

\bibitem[{{Petrosian}(1976)}]{1976ApJ...209L...1P}
{Petrosian}, V. 1976, \apjl, 209, L1

\bibitem[{{Ry{\'s}} {et~al.}(2014){Ry{\'s}}, {van de Ven}, \&
  {Falc{\'o}n-Barroso}}]{Rys2014}
{Ry{\'s}}, A., {van de Ven}, G., \& {Falc{\'o}n-Barroso}, J. 2014, \mnras, 439,
  284

\bibitem[{{S{\'a}nchez-Janssen} {et~al.}(2010){S{\'a}nchez-Janssen},
  {M{\'e}ndez-Abreu}, \& {Aguerri}}]{SanchezJanssen2010}
{S{\'a}nchez-Janssen}, R., {M{\'e}ndez-Abreu}, J., \& {Aguerri}, J.~A.~L. 2010,
  \mnras, 406, L65

\bibitem[{{Sandage}(1986)}]{1986A&A...161...89S}
{Sandage}, A. 1986, \aap, 161, 89

\bibitem[{{Shu}(1969)}]{1969ApJ...158..505S}
{Shu}, F.~H. 1969, \apj, 158, 505

\bibitem[{{Smith} {et~al.}(2010){Smith}, {Davies}, \&
  {Nelson}}]{2010MNRAS.405.1723S}
{Smith}, R., {Davies}, J.~I., \& {Nelson}, A.~H. 2010, \mnras, 405, 1723

\bibitem[{{Smith} {et~al.}(2012){Smith}, {Lucey}, {Price}, {Hudson}, \&
  {Phillipps}}]{SmithRussell2012}
{Smith}, R.~J., {Lucey}, J.~R., {Price}, J., {Hudson}, M.~J., \& {Phillipps},
  S. 2012, \mnras, 419, 3167

\bibitem[{{Spitzer}(1958)}]{1958ApJ...127...17S}
{Spitzer}, Jr., L. 1958, \apj, 127, 17

\bibitem[{{Springel} {et~al.}(2005){Springel}, {White}, {Jenkins}, {Frenk},
  {Yoshida}, {Gao}, {Navarro}, {Thacker}, {Croton}, {Helly}, {Peacock}, {Cole},
  {Thomas}, {Couchman}, {Evrard}, {Colberg}, \& {Pearce}}]{2005Natur.435..629S}
{Springel}, V., {White}, S.~D.~M., {Jenkins}, A., {et~al.} 2005, \nat, 435, 629

\bibitem[{{Spurzem}(1999)}]{1999JCoAM.109..407S}
{Spurzem}, R. 1999, Journal of Computational and Applied Mathematics, 109, 407

\bibitem[{{Spurzem} {et~al.}(2008){Spurzem}, {Berentzen}, {Berczik}, {Merritt},
  {Amaro-Seoane}, {Harfst}, \& {Gualandris}}]{Spurzem2008}
{Spurzem}, R., {Berentzen}, I., {Berczik}, P., {et~al.} 2008, in Lecture Notes
  in Physics, Berlin Springer Verlag, Vol. 760, The Cambridge N-Body Lectures,
  ed. S.~J. {Aarseth}, C.~A. {Tout}, \& R.~A. {Mardling}, 377

\bibitem[{{Spurzem} {et~al.}(2009){Spurzem}, {Giersz}, {Heggie}, \&
  {Lin}}]{2009ApJ...697..458S}
{Spurzem}, R., {Giersz}, M., {Heggie}, D.~C., \& {Lin}, D.~N.~C. 2009, \apj,
  697, 458

\bibitem[{{Toloba} {et~al.}(2011){Toloba}, {Boselli}, {Cenarro}, {Peletier},
  {Gorgas}, {Gil de Paz}, \& {Mu{\~n}oz-Mateos}}]{2011A&A...526A.114T}
{Toloba}, E., {Boselli}, A., {Cenarro}, A.~J., {et~al.} 2011, \aap, 526, A114

\bibitem[{{Toloba} {et~al.}(2014){Toloba}, {Guhathakurta}, {Boselli},
  {Peletier}, {Emsellem}, {Lisker}, {van de Ven}, {Simon}, {Falcon-Barroso},
  {Adams}, {Benson}, {Boissier}, {den Brok}, {Gorgas}, {Hensler}, {Janz},
  {Laurikainen}, {Paudel}, {Rys}, \& {Salo}}]{Toloba2014p3}
{Toloba}, E., {Guhathakurta}, P., {Boselli}, A., {et~al.} 2014, \apj, in press

\bibitem[{{Tully} \& {Trentham}(2008)}]{2008AJ....135.1488T}
{Tully}, R.~B. \& {Trentham}, N. 2008, \aj, 135, 1488

\bibitem[{{Velazquez} \& {White}(1999)}]{1999MNRAS.304..254V}
{Velazquez}, H. \& {White}, S.~D.~M. 1999, \mnras, 304, 254

\bibitem[{{Villalobos} {et~al.}(2012){Villalobos}, {.}, {De Lucia}, {Borgani},
  \& {Murante}}]{2012MNRAS.424.2401V}
{Villalobos}, {\'A}., {.}, {De Lucia}, G., {Borgani}, S., \& {Murante}, G.
  2012, \mnras, 424, 2401

\bibitem[{{Weinmann} {et~al.}(2010){Weinmann}, {Kauffmann}, {von der Linden},
  \& {De Lucia}}]{2010MNRAS.406.2249W}
{Weinmann}, S.~M., {Kauffmann}, G., {von der Linden}, A., \& {De Lucia}, G.
  2010, \mnras, 406, 2249

\bibitem[{{Wetzel} {et~al.}(2013){Wetzel}, {Tinker}, {Conroy}, \& {van den
  Bosch}}]{2013MNRAS.432..336W}
{Wetzel}, A.~R., {Tinker}, J.~L., {Conroy}, C., \& {van den Bosch}, F.~C. 2013,
  \mnras, 432, 336

\end{thebibliography}

\end{document}